\newif\ifsingle

\newif\ifFullVersion
\FullVersiontrue 

\ifsingle
\documentclass[11pt,draftclsnofoot, onecolumn]{IEEEtran}		
\else		
\documentclass[10pt,final, twocolumn]{IEEEtran}
\fi

\usepackage{times}
\usepackage{amsmath,dsfont}
\usepackage{amssymb,amsthm}
\usepackage{epsfig,verbatim}
\usepackage{subfigure}
\usepackage{setspace}
\usepackage{color}
\usepackage{cite}
\usepackage{epstopdf}
\usepackage{graphics}
\usepackage{accents}
\usepackage{acronym}
\usepackage[bookmarks,colorlinks]{hyperref}
\usepackage{booktabs}
\usepackage{mathtools}
\usepackage{enumitem}
\usepackage{amsmath}
\DeclareMathOperator*{\argmax}{argmax}
\usepackage{multirow}

\usepackage[ruled,linesnumbered]{algorithm2e}  
\SetKwInput{KwData}{\textbf{Init}} 

\newcommand{\myVec}[1]{{\boldsymbol{#1}}}
\newcommand{\myMat}[1]{{\boldsymbol{#1}}}
\newcommand{\mySet}[1]{\mathcal{#1}}
\newcommand{\E}{\mathds{E}}		 			
\newcommand{\myI}{{\myMat{I}}}			 		
\newcommand{\myX}{{\myVec{x}}}			 		
\newcommand{\myY}{\myVec{y}}
\newcommand{\myZ}{\myVec{z}}
\newcommand{\myS}{\myVec{s}}

\newcommand{\myH}{\myMat{H}}
\newcommand{\lenX}{n}			 			
\newcommand{\lenS}{k}			 			
\newcommand{\Pdf}[1]{f_{ #1}}
\newcommand{\lenZ}{p}			 			
\newcommand{\Qres}{\tilde{M}}
\newcommand{\SmpSize}{\tilde{L}}
\newcommand{\Basis}{B}
\newcommand{\NyqSize}{L}
\newcommand{\Period}{P}
\newcommand{\Interval}{T}
\newcommand{\NyqInt}{T_{\NyqSize}}
\newcommand{\Quan}{Q_{\Qres}}
\newcommand{\SampTr}{\phi_{\SmpSize}}
\newcommand{\QuanTr}{q_{\Qres}}
\newcommand{\Samp}{\Phi_{\SmpSize}}
\newcommand{\Hparams}{\myVec{\psi}}
\newcommand{\Acqparams}{\myVec{\theta}}
\newcommand{\NetMap}{\gamma_{\Hparams,\Acqparams}}
\newcommand{\BaseF}{u}

\newcommand{\BitBudget}{\mathcal{B}}

\newcommand{\Ntraining}{N}

\newtheorem{example}{Example}

\ifsingle

\newcommand{\includefig}[1]{\includegraphics[width = 0.75\columnwidth]{#1} 	\vspace{-0.2cm}}
\else

\newcommand{\includefig}[1]{\includegraphics[width = \columnwidth]{#1} 	\vspace{-0.4cm}}

\fi 

\acrodef{adc}[ADC]{analog-to-digital convertor}  
\acrodef{csi}[CSI]{channel state information} 
\acrodef{snr}[SNR]{signal-to-noise ratio}
\acrodef{bs}[BS]{base station}  
\acrodef{mimo}[MIMO]{multiple-input multiple-output}
\acrodef{mse}[MSE]{mean-squared error}
\acrodef{pdf}[PDF]{probability density function}
\acrodef{rv}[RV]{random variable} 
\acrodef{isi}[ISI]{intersymbol interference}  
\acrodef{awgn}[AWGN]{additive white Gaussian noise} 
\acrodef{lti}[LTI]{linear time-invariant}  
\acrodef{ut}[UT]{user terminal} 
\acrodef{mmw}[mmWave]{millimeter wave}
\acrodef{dma}[DMA]{dynamic metasurface antenna}
\acrodef{ofdm}[OFDM]{orthogonal frequency division multiplexing}
\acrodef{dnn}[DNN]{deep neural network}
\acrodef{ml}[ML]{machine learning}
\acrodef{sgd}[SGD]{stochastic gradient descent}
\acrodef{bpsk}[BPSK]{binary phase shift keying}
\acrodef{ber}[BER]{bit error rate}
\acrodef{csi}[CSI]{channel state information}
\acrodef{map}[MAP]{maximum a-posteriori probability}
\acrodef{ct}[CT]{continuous-time}
\acrodef{fri}[FRI]{finite rate of innovation}

\IEEEoverridecommandlockouts

\title{Deep Task-Based Analog-to-Digital Conversion
}
\author{
	\IEEEauthorblockN{Nir Shlezinger, Ariel Amar, Ben Luijten, Ruud J. G. van Sloun, and Yonina C. Eldar\\
	} 
	\thanks{
	    Parts of this work were presented in the IEEE International Conference on Acoustics, Speech, and Signal Processing (ICASSP) 2020 as the paper \cite{shlezinger2020learning}.
		This project has received funding from the European Union’s Horizon 2020 research and innovation program under grant No. 646804-ERC-COG-BNYQ, and from the Israel Science Foundation under grant No. 0100101.
        N. Shlezinger, is with the School of ECE, Ben-Gurion University of the Negev, Beer-Sheva, Israel (e-mail: nirshl@bgu.ac.il).
A. Amar  and Y. C. Eldar are with the Faculty of Math and CS, Weizmann Institute of Science, Rehovot, Israel (e-mail: arielamar123@gmail.com; yonina.eldar@weizmann.ac.il).
		B. Luitjen and R. J. G. van Sloun are with the EE Dpt., Eindhoven University of Technology,  The Netherlands (e-mail:\{w.m.b.luijten; r.j.g.v.sloun\}@tue.nl). R. J. G. van Sloun is also with Phillips Research, Eindohoven,  The Netherlands.
	}

	\vspace{-1.0cm}
	
}
\vspace{-0.75cm}

\begin{document}
	
	\maketitle
	\pagestyle{plain}
	\thispagestyle{plain}
	\begin{abstract}
		Analog-to-digital converters (ADCs) allow physical signals to be processed using digital hardware. Their conversion consists of two stages: Sampling, which maps  a continuous-time signal into discrete-time, and quantization, i.e., representing the continuous-amplitude quantities using a finite number of bits. ADCs typically implement generic uniform conversion mappings that are ignorant of the task for which the signal is acquired, and can be costly when operating in high rates and fine resolutions.  
		In this work we design task-oriented ADCs which learn from data how to map an analog signal into a  digital representation such that the system task can be efficiently carried out. We propose a model for sampling and quantization that facilitates the learning of   non-uniform mappings from  data. Based on this learnable ADC mapping, we present a mechanism for  optimizing a hybrid acquisition system comprised of analog combining, tunable ADCs with fixed rates, and digital processing, by jointly learning its components  end-to-end. Then, we show how one can exploit the representation of hybrid acquisition systems as deep network to optimize the sampling rate and quantization rate given the task by utilizing Bayesian meta-learning techniques.
		We evaluate the proposed deep task-based ADC in two case studies: the first considers symbol detection in multi-antenna digital receivers, where multiple analog signals are simultaneously acquired in order to recover a set of discrete information symbols. The second application is the beamforming of analog channel data acquired in ultrasound imaging. Our numerical results demonstrate that the proposed approach achieves performance which is comparable to operating with high sampling rates and fine resolution quantization, while operating with reduced overall bit rate. For instance, we demonstrate that deep task-based ADCs enable accurate reconstruction of ultrasound images while using $12.5\%$ of the overall number of bits used by conventional ADCs to achieve similar performance. 
		
	\end{abstract}
	
	\vspace{-0.4cm}
	\section{Introduction}
	\vspace{-0.1cm} 
	A multitude of electronic systems process physical signals using digital hardware. Digital signal processors  represent analog quantities as a set of bits using analog-to-digital conversion. 
	Converting a \ac{ct} signal taking continuous-amplitude values into a finite-bit representation consists of two steps: The analog signal is first sampled into a discrete-time process, which is then quantized into discrete-amplitude values, such that it can be digitally processed~\cite{eldar2015sampling}. 
	
	The acquisition of analog signals is commonly carried out using scalar \acp{adc} \cite{walden1999analog}.  These devices sample the \ac{ct} signal in uniformly spaced time-instances and obtain a digital representation using a uniform mapping of the real line.  While this acquisition strategy is simple to implement, it is limited in its ability to accurately represent   signals in digital \cite{kipnis2018analog}, especially when operating under constrained sampling rate and low quantization resolution, due to, e.g., cost, power, or memory constraints. 
	Furthermore, this procedure is carried out regardless of the task for which the analog signal is acquired into a digital representation. 
	
	In practice, analog signals are often acquired in order to extract some underlying information, namely, for a task other than recovering the analog process. One example is \ac{mimo} communications receivers, which  recover a transmitted discrete message from their observed channel output. \ac{mimo} receivers typically operate under strict power and cost constraints, which are particularly relevant when operating in high frequency bands \cite{xiao2017millimeter}.  Another relevant example is ultrasound imaging, where  large amounts of analog channel data is acquired to form an image, which notably affects the hardware cost and complexity \cite{chernyakova2014fourier}.   While designing energy efficient uniform \acp{adc} is an on-going area of research \cite{yazicigil2019taking,jain2020esampling,6936944,8727467}, a natural approach to relieve the harmful affects of high resolution acquisition is to restrict the sampling rate and quantization resolution of the \acp{adc}.
	
	When acquiring for a specific task, it was  shown in \cite{shlezinger2018hardware,shlezinger2018asymptotic,Salamtian19task,shlezinger2019deep,huijben2019learning, mulleti2021learning, solodky2018sampling,liu2018analog,neuhaus2020task} that the distortion induced by sample and bit limitations can be notably reduced by accounting for the task in  acquisition. In particular, the works \cite{shlezinger2018hardware,shlezinger2018asymptotic,Salamtian19task} analytically designed task-based quantization systems for estimation tasks  by introducing analog processing and tuning the quantization rule, assuming ideal (Nyquist rate) samplers;  A data-driven approach for designing task-based quantizers under generic setups was considered in \cite{shlezinger2019deep}, which utilized \ac{ml} tools; The works \cite{huijben2019learning,mulleti2021learning} also used \ac{ml} to learn sampling mechanisms  assuming error-free quantization, while \cite{solodky2018sampling} and \cite{liu2018analog} analytically designed samplers for maximizing capacity in \ac{mimo} systems and for audio classification, respectively. However, none of these  works study the full acquisition process. For acquisition involving both sampling and quantization, the work \cite{neuhaus2020task} studied the analytical design of hybrid analog/digital acquisition systems with uniform \acp{adc} for recovering linear functions of their observations. Furthermore, the analysis of joint sampling and quantization systems was considered in \cite{kipnis2016distortion,kipnis2018fundamental}, which focused on complex  source coding instead of scalar quantization, and derived bounds on the reconstruction accuracy in the absence of a task. The design of acquisition systems with possibly non-uniform ADCs for a  (possibly analytically intractable) task, has not yet been studied, and is the focus here. 
	
	In this work we propose a task-based acquisition system utilizing scalar \acp{adc} for signals obeying a finite basis expansion model. As analytically deriving task-based methods  is difficult and commonly requires imposing a limited structure,  
such as assuming  uniform \ac{adc} mappings and linear operations 
\cite{shlezinger2018hardware,Salamtian19task,neuhaus2020task},  we adopt a data-driven approach based on \ac{ml}. We  design the system to learn its sampling and quantization mappings along with its analog and digital processing from training data, such that it can reliably carry out its task. 
A major challenge in designing \ac{ml}-based \acp{adc} and incorporating such devices into \acp{dnn}, stems from the continuous-to-discrete nature of sampling and quantization: These operations are either non-differentiable or nullify the gradient \cite{shlezinger2019deep,huijben2019learning},  limiting the application of conventional training based on backpropagation. To overcome this, we adopt a soft-to-hard approach based on gradient estimation through relaxation of the discrete process \cite{agustsson2017soft}, also utilized in \cite{shlezinger2019deep} for optimizing quantization mappings. We propose a differentiable approximation of sampling, which can be trained to learn non-uniform sampling methods, and is combined with the trainable quantizer of \cite{shlezinger2019deep} into a dynamic data-driven \ac{adc}. We incorporate this adaptive \ac{adc} into a \ac{dnn} architecture resulting in a deep task-based acquisition system, which can be trained using conventional training methods, e.g., \ac{sgd} with backpropagation. 

By representing the hybrid analog/digital acquistion system as a trainable \ac{dnn} with non-conventional layers encapsulating the sampling and quantization operations, we are also able to optimize the parameters dictating the overall bit rate. By treating the sampling rate, the quantization resolution, and the number of scalar \acp{adc}, as hyperparameters of the \ac{dnn}, we propose a meta-learning scheme to optimize these key quantities from data. We utilize a variation of Bayesian optimization based meta-learning \cite{frazier2018tutorial}  in order to minimize these key quantities while preserving the ability to accurately carry out the task.
	Our proposed deep task-based hybrid acquisition system is evaluated in two applications: A synthetic  scenario of detection from linear observations, and an ultrasound beamforming setup. For the first application,   we demonstrate the ability of deep task-based \acp{adc} to achieve comparable performance to the  \ac{map} estimator {\em without quantization constraints}, and to outperform the common approach of processing only in the digital domain with uniform \acp{adc}. Our proposed meta-learning scheme is shown to notably reduce the overall bit rate  without degrading the performance, and in some cases even improving the detection accuracy. For ultrasound beamforming, we show how the conversion of analog channel data into a set of pixels specializes a task-based acquisition setup. Then, we demonstrate that deep task-based \acp{adc} can use as low as $12.5\%$ the number of bits compared to conventional acquisition while hardly affecting the quality of the recovered  image.  
	
	The rest of this paper is organized as follows: 
Section~\ref{sec:Model}  presents the  system model.
Section~\ref{sec:ADCs} details the proposed  deep task-based \ac{adc} system, while Section~\ref{sec:Meta} presents the Bayesian meta-learning scheme for optimizing its configuration. The synthetic case study and the application to ultrasound beamforming are discussed in Sections~\ref{sec:AppMIMO} and \ref{sec:AppBF}, respectively. Finally,  Section~\ref{sec:Conclusions}  concludes the paper.

	Throughout the paper, we use boldface lower-case letters for vectors, e.g., ${\myVec{x}}$;
	the $i$th element of ${\myVec{x}}$ is written as $({\myVec{x}})_i$. 
	Boldface upper-case letters denote matrices,  e.g., 
	$\myMat{M}$; and  $(\myMat{M})_{i,j}$   is its $(i,j)$th element. 
	%
	%
	Finally, $\mySet{R}$ and $\mySet{Z}$,  ${\rm sign}(\cdot)$ and $\delta(\cdot)$ are the sets of real numbers, integers, sign, and Dirac delta function,  respectively.

	\vspace{-0.2cm}
	\section{Task-Based Signal Acquisition Setup}
	\label{sec:Model} 
	\vspace{-0.1cm}
	In this section we formulate the task-based acquisition setup. 
	We first present the system model in Subsection~\ref{subsec:Model_Acquisition}, after which we formulate the problem of designing such systems in a data-driven manner in Subsection~\ref{subsec:Model_Problem}. 
	 	
	\vspace{-0.1cm}
	\subsection{System Model}
	\label{subsec:Model_Acquisition}
\begin{figure*}
		\centering
		\includegraphics[width=\linewidth]{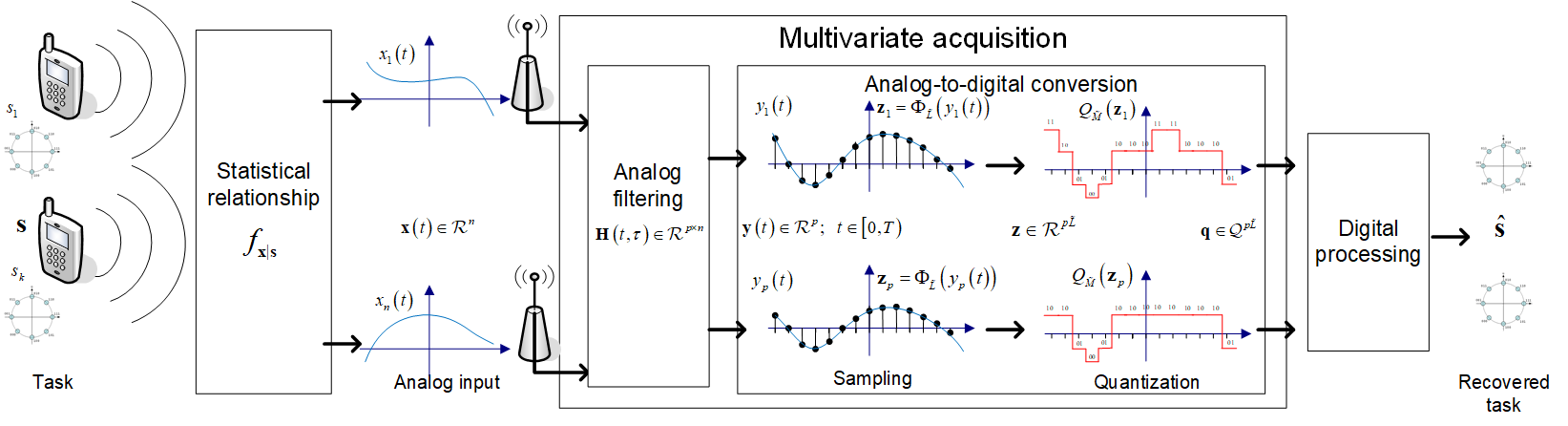}
		\vspace{-0.2cm}
		\caption{Hybrid task-based acquisition system illustration. The task here is recovering a set of constellation symbols in  \ac{mimo} communications. }
		\ifsingle
		\vspace{-0.2cm}
		\else	
		\vspace{-0.45cm}
		\fi
		\label{fig:SystemModel}
	\end{figure*}

	\vspace{-0.1cm}	 
	The task-based acquisition setup is modeled using the hybrid system illustrated in Fig. \ref{fig:SystemModel}. The system consists of analog filtering, analog-to-digital conversion, and digital processing. Our goal, as detailed in Subsection~\ref{subsec:Model_Problem}, is to propose a mechanism for learning these components from data. We focus on scenarios where a set of $\lenX$ analog signals $\{x_i(t)\}_{i=1}^{\lenX}$  are  converted into a digital representation in order to recover an unknown vector $\myS\in\mySet{S}^{\lenS}$, referred to as {\em the system task}. The task vector $\myS$ is statistically related to the multivariate analog signal $\myX(t) \triangleq [x_1(t), \ldots x_\lenX(t)]^T$ via a conditional distribution $\Pdf{\myX|\myS}$. Such scenarios represent, for example, measurements taken from sensor arrays  to detect some physical phenomenon, acoustics echos acquired in order to form an image in ultrasound beamforming, or channel outputs acquired by a \ac{mimo} receiver  for decoding a transmitted message.

	{\bf Signal Model:} We focus on the case where each of the \ac{ct} signals known to be spanned by a set of $\Basis$ basis functions, i.e., for each $x_i(t)$ there exists a set of functions $\{\BaseF_{i,j}(t)\}_{j=0}^{\Basis-1}$ and coefficients $\{\tilde{x}_{i,j}\}_{j=0}^{\Basis-1}$ such that $x_i(t) = \sum_{j=0}^{\Basis-1} \tilde{x}_{i,j}\BaseF_{i,j}(t)$. 
	By defining the $\lenX \times \lenX$ diagonal matrix $\myMat{U}_j(t)$ such that $(\myMat{U}_j(t))_{i,i}(t) \triangleq \BaseF_{i,j}(t)$ and the vector $\tilde{\myX}_j \triangleq [ \tilde{x}_{0,j}, \ldots,  \tilde{x}_{\lenX-1,j}]^T$, we can write the multivariate signal as 
		\begin{equation}
	\label{eqn:XModel2}
	 \myX(t) = \sum_{j=0}^{\Basis-1} \myMat{U}_j(t) \tilde{\myX}_j.  
	\end{equation}
	The generic formulation in \eqref{eqn:XModel2} accommodates a broad family of signals, as exemplified next:
	\begin{example}
	\label{exm:FRI}
	    Let $\myX(t)$ be a set of \ac{fri} signals, i.e., there exists a function $u(t)$ and a set $\{\tau_{i,j}\}$ such that $x_i(t) = \sum_{j=0}^{\Basis-1} \tilde{x}_{i,j}u(t-\tau_{i,j})$ \cite{mulleti2021learning}. Such signals are a special case of \eqref{eqn:XModel2}, obtained by setting  $u_{i,j}(t) = u(t-\tau_{i,j})$.
	\end{example}
    \begin{example}
    \label{exm:FiniteSupprt}
        Let $\myX(t)$ be a set of signals defined over a finite time interval $t\in[0,\Period)$ for some $\Period>0$, and thus each $x_i(t)$ can be written using its Fourier series expansion. Such signals with finite Fourier series are a special case of \eqref{eqn:XModel2} obtained when $u_{i,j}(t)$ is the $j$th Fourier basis function for $t\in[0,\Period)$. 
    \end{example}
    \begin{example}
    \label{exm:Periodic}
        When $\myX(t)$ is periodic  with period $\Period>0$ and has a finite Fourier series expansion, then it can be written via  \eqref{eqn:XModel2} with $u_{i,j}(t)$ being the $j$th Fourier basis function $\forall t \in \mySet{R}$.
    \end{example}
	
	The motivation for this model is that it allows us to rigorously express the multivariate \ac{ct} signal $\myX(t)$ using the $\lenX\Basis\times 1$ vector $\myX \triangleq [\tilde{\myX}_0^T, \ldots, \tilde{\myX}_{\Basis-1}^T ]^T$, which encapsulates the information needed to recover the task vector $\myS$.
	
	{\bf Analog Filtering:} The observed $\myX(t)$ is first mapped into a set of $\lenZ$ \ac{ct} signals  $\{y_i(t)\}_{i=1}^{\lenZ}$, representing the processing carried out in analog. We focus on linear analog processing, in which  $\myY(t) \triangleq [y_1(t), \ldots y_\lenZ(t)]^T$ is obtained from $\myX(t)$ via multivariate filtering with a matrix impulse response $\myH(t,\tau)\in \mySet{R}^{\lenZ\times\lenX}$, i.e., 
\begin{equation}
    \label{eqn:AnalogFilt}
\myY(t) = \int \myH(t,t-\tau) \myX(\tau) d\tau.
\end{equation} 
We do not restrict the  filter to be \ac{lti} or causal, allowing it to represent a broad range of acquisition systems. 
We focus on linear operation being a common model for feasible analog processing;  linear analog filters were shown to facilitate the sampling and recovery of multivariate  signals in \cite{shlezinger19joint}, while \ac{lti} memoryless combiners are commonly used for RF chains reduction in \ac{mimo} systems \cite{mendez2016hybrid,ioushua2019family,gong2019rf}. 

Based on the signal model in \eqref{eqn:XModel2}, it holds that for each $t \in \mySet{R}$, the vector $\myY(t)$ is given by a linear function of $\myX$, as
\begin{align}
\myY(t) &= \int \myH(t,t-\tau)  \sum_{j=0}^{\Basis-1} \myMat{U}_j(\tau) \tilde{\myX}_j d\tau \notag \\
&=  \sum_{j=0}^{\Basis-1} \left(\int \myH(t,t-\tau)\myMat{U}_j(\tau) d\tau \right)\tilde{\myX}_j = \tilde{\myMat{H}}(t) \myX, 
\label{eqn:NewH}
\end{align}
where $\tilde{\myMat{H}}(t)$ is an $\lenZ \times \lenX\Basis$ block-matrix comprised of a row of $\Basis$ sub-matrices, i.e., $\tilde{\myMat{H}}(t) = [\tilde{\myMat{H}}_0(t), \ldots, \tilde{\myMat{H}}_{\Basis-1}(t)]$ where the $j$th submatrix is 
\begin{equation}
    \label{eqn:NewH2}
   \tilde{\myMat{H}}_j(t) \triangleq \int \myH(t,t-\tau)\myMat{U}_j(\tau) d\tau.
\end{equation}
\begin{example}
    \label{exm:LTIPeriodic}
    When $\myX(t)$ is periodic as in Example~\ref{exm:Periodic} and the analog filter is \ac{lti}, then $\tilde{\myMat{H}}_j(t) = \mathcal{F}\{\myMat{H}\}\big(\frac{2\pi j}{\Period}\big)\myMat{U}_j(t) $,  where $\mathcal{F}\{\myMat{H}\}(\omega)$ is the multivariate frequency response of the filter, and $\myMat{U}_j(t)$ is the $j$th Fourier basis as in Example~\ref{exm:Periodic}.
    \end{example}

	{\bf Analog-to-Digital Conversion: }
	Next, $\myY(t)$ observed over the interval $t\in [0,\Interval)$ is converted into a digital representation using a set of scalar \acp{adc}. 
	Each signal $y_i(t)$ undergoes the same \ac{adc} mapping, which consists of arbitrary non-uniform sampling and quantization: 
	Sampling is represented by the operator $\Samp(\cdot)$, such that $\myZ_i =  \Samp(y_i(t))$ is a $\SmpSize\times 1$ vector whose entries are $(\myZ_i)_j = y_i(t_j)$, where $\{t_j\}_{j=1}^{\SmpSize} \subset [0,\Interval)$, i.e.,
	\begin{equation}
	\label{eqn:Sampler}
	\left( \Samp(\alpha(t))\right)_j = \int \alpha(\tau) \delta(\tau - t_j)d\tau, \qquad j\in \{1,\ldots,\SmpSize\}.
	\end{equation} 
	The parameters $\{t_j\}_{j=1}^{\SmpSize}$ determine the sampling times, which are not restricted to represent uniform sampling. 
	
	Quantization is carried out using a continuous-to-discrete mapping $\Quan:\mySet{R}\mapsto \mySet{Q}$ applied to each entry of $\myZ_i$, where $\Qres = |\mySet{Q}|$ is the  resolution, i.e., it uses $\lceil\log_2 \Qres\rceil$  bits. The mapping is given by (almost everywhere on $\mySet{R}$)
	\begin{equation}
	\label{eqn:quantizer}
	\Quan(\alpha) = a_0 + \sum_{i=1}^{\Qres-1}a_i {\rm sign}\left( \alpha - b_i\right).
	\end{equation} 	
	In \eqref{eqn:quantizer},
	$\{b_i\}$ and $\{a_i\}$ determine the decision regions and their assigned values, i.e., the set $\mySet{Q}$. For the special case of uniform quantization, the difference $b_{i}-b_{i-1}$ and the values of $a_i$ are constant, and do not depend on the decision region index $i$.
	By defining $\myZ \triangleq [\myZ_1^T,\ldots, \myZ_{\lenZ}^T]^T$, the output of the \acp{adc} is the vector $\myVec{q} \in \mySet{Q}^{\lenZ\SmpSize}$ whose entries are $(\myVec{q})_l = \Quan((\myZ)_l)$. The overall number of bits used for acquisition is   $\lenZ \cdot \SmpSize \cdot \lceil\log_2 \Qres\rceil$. 
	
	{\bf Digital Processing: }
	The discrete  vector $\myVec{q}$ is processed in  digital to estimate  the  task vector $\myS$ as $\hat{\myS} \in \mySet{S}^{\lenS}$.

	\vspace{-0.1cm}
	\subsection{Problem Formulation}
	\label{subsec:Model_Problem}
	\vspace{-0.1cm}			
	Designing task-based acquisition systems using model-based methods, namely, analytically setting the filter $\myH(t,\tau)$ and the operators $\Samp(\cdot)$ and $\Quan(\cdot)$ based on $\Pdf{\myX|\myS}$, is very difficult. Consequently, previous model-based studies  assumed  some specific conditional distribution $\Pdf{\myX|\myS}$ with either fixed sampling rule and uniform quantizers \cite{shlezinger2018hardware,Salamtian19task, neuhaus2020task}, or alternatively, considering error-free quantization \cite{solodky2018sampling}. Furthermore, accurate knowledge of  $\Pdf{\myX|\myS}$ may not be available in practice. Consequently, our goal is to design   task-based acquisition systems in a data-driven fashion using \ac{ml} methods. 
	
	To formulate the setup such that it can be designed using \ac{ml} tools,  which typically operate on vectors and not on \ac{ct} quantities, we  restrict the sampling times $\{t_j\}$, which dictate $\Samp(\cdot)$, to be a subset of the some dense uniform grid. We divide the observation interval $[0,\Interval)$ into $\NyqSize$ sub-intervals of duration $\NyqInt = \frac{\Interval}{\NyqSize}$, and select our sampling points from the set $\{l\NyqInt\}_{l=0}^{\NyqSize-1}$. The grid is set such that the number of samples taken by the acquisition system $\SmpSize$ is smaller, and preferably much smaller, than the size of the dense grid $\NyqSize$. Here, by writing the dense samples as the vector $\myY \triangleq [\myY^T(0), \ldots, \myY^T((\NyqSize-1)\NyqInt) ]^T$, it holds that the input to the quantizer $\myZ$ consists of entries of $\myY$.  
	Constraining the acquisition system to sample from a discretized grid facilitates its design using \ac{ml} methods, as done in \cite{huijben2019learning}. Furthermore, by defining the $\NyqSize \lenZ \times \lenX \Basis$ matrix $\bar{\myMat{H}}$ which is comprised of a column of $\NyqSize$ sub-matrices with $\tilde{\myMat{H}}(l \NyqInt)$ being the $l$th sub-matrix, it follows from \eqref{eqn:NewH} that the candidate samples are expressed as
	\begin{equation}
	\label{eqn:SamplingGrid}
	    \myY = \bar{\myMat{H}} \myX.
	\end{equation}
	
	It is noted that the matrix may be restricted to take a given structure, depending on the constrains imposed on the filter $\myMat{H}(t, \tau)$ as well as the specification of the basis functions in \eqref{eqn:XModel2}. Accordingly, once such a structure is imposed, one can trace $\bar{\myMat{H}}$ to the setting of the \ac{ct} filter $\myMat{H}(t, \tau)$. To how such structures are obtained, consider the following example:
	\begin{example}\label{exm:BlockTop}
	    Consider \ac{fri} signals where  the delays lie on the sampling grid, such that $\myX(t)$ in Example~\ref{exm:FRI} can be written with $\tau_{i,j} = j\NyqInt $ and $\Basis = \NyqSize$. When the analog filter is \ac{lti}, i.e.,  $\myMat{H}(t,\tau) \equiv \myMat{H}(\tau)$, it holds by \eqref{eqn:NewH2} that 
	    \begin{align}
	        \tilde{\myMat{H}}_j(i\NyqInt) &= \int \myMat{H}(i\NyqInt-\tau) \BaseF(\tau -j\NyqInt) d\tau \notag \\
	        &= \myMat{G}((i-j)\NyqInt),
	    \end{align}
	    where we define $\myMat{G}(t)$ as the convolution $ \int \myMat{H}(t-\tau) \BaseF(\tau) d\tau$. Under the given constraints, $\bar{\myMat{H}}$ is a block-Toeplitz matrix.
	\end{example}

	The task-based acquisition system is thus required to learn to recover $\myS$ from $\myX(t)$ based on a training set $\{\myS^{(j)}, \myX^{(j)}\}_{j=1}^{\Ntraining}$ consisting of $\Ntraining$ realizations of the densely-sampled inputs and their corresponding task vectors. In particular, the system parameters, i.e., the analog filter $ \bar{\myMat{H}}$, sampling operator $\Samp(\cdot)$, quantization rule $\Quan(\cdot)$, and the processing of the digital vector $\myVec{q}$ into  $\hat{\myS}$, are learned from training. The hybrid acquisition system is restricted to utilize at most  $\BitBudget$ bits i.e.,  $\lenZ \cdot \SmpSize \cdot \lceil\log_2 \Qres\rceil \leq \BitBudget$. 
	Once these parameters are tuned, they can be configured into the task-based acquisition system detailed in the previous subsection, which operates on \ac{ct} analog signals.

	\vspace{-0.2cm}
	\section{Learning Task-Based Acquisition}
	\label{sec:ADCs}
	\vspace{-0.1cm} 
	In this section we present our deep task-based acquisition system, which learns how to map  analog signals into an estimate of the task vector. We focus here on optimizing the hybrid acquisition system when number of \acp{adc} $\lenZ$, the amount of samples acquired $\SmpSize$, and the quantizer resolution $\Qres$, are fixed, and embodied in the hyperparemeter vector $\Acqparams = [\lenZ, \SmpSize, \log_2 \Qres]$. The optimization of these parameter subject to an overall bit constraint $\BitBudget$ is discussed in Section~\ref{sec:Meta}.
	We begin by detailing the analog and digital networks in Subsection \ref{subsec:Architecture}. Then, in Subsection~\ref{subsec:Layers} we present how \ac{adc} mappings are trained, and discuss the resulting structure in Subsection~\ref{subsec:Discussion}. Throughout this section we consider generic tasks and \ac{dnn} architectures. Specific models are detailed and  evaluated   in the case studies presented in  Sections~\ref{sec:AppMIMO}-\ref{sec:AppBF}. 

	\vspace{-0.2cm}
	\subsection{Learned Analog and Digital Processing}
	\label{subsec:Architecture}
\begin{figure}
		\centering
		{\includefig{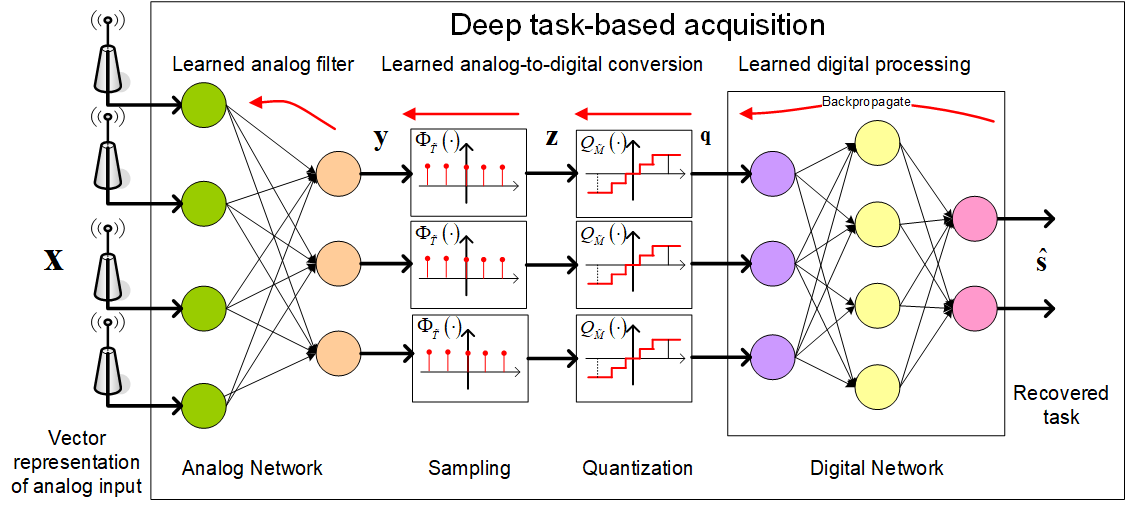}} 
		\caption{Deep task-based analog-to-digital conversion illustration.}
		\label{fig:GenSetup2}	 
	\end{figure}

	\vspace{-0.1cm}
	Our proposed deep task-based acquisition system implements the analog filtering and digital processing 
	using dedicated \acp{dnn}, denoted as the {\em analog network} and the {\em digital network}, respectively. An illustration of such a system is depicted in Fig. \ref{fig:GenSetup2}. 
	The input to the system is the vector representation of the observed signal $\myX$. 
	To realize linear filters, as detailed in our system model in Section \ref{sec:Model}, 
	the analog network should consist only of linear layers, while  introducing  non-linear activations yields non-linear analog processing.

	While the network architecture illustrated in Fig.~ \ref{fig:GenSetup2} is generic, the recovery of the vector $\myS$ of interest can be broadly divided into two types of tasks: classification and estimation (regression).
	When $\mySet{S}$ is a finite set, which is the case in the \ac{mimo} detection application presented in Section~\ref{sec:AppMIMO} the recovery of $\myS$ can be viewed as classifying from $|\mySet{S}|^{\lenS}$ possible categories. In such cases, the output layer of the digital network is a softmax layer with $|\mySet{S}|^{\lenS}$ outputs, each representing the conditional distribution of the corresponding  label  given the input. The overall network is trained  end-to-end  to minimize the  cross-entropy loss. 
	By letting $\Hparams$ be the set of system trainable parameters and $\NetMap(\myX; \myVec{\alpha})$ be the output corresponding to $\myVec{\alpha}\in\mySet{S}^\lenS$ with network weights $\Hparams$ and \ac{adc} hyperparameters $\Acqparams$,  the loss function is given by
	\vspace{-0.1cm}
	\begin{equation}
	\label{eqn:LossFuncCE}
	\mathcal{L}_{\Acqparams}(\Hparams)  =\frac{1}{\Ntraining}\sum_{j=1}^{\Ntraining} -\log  \NetMap\Big(\myVec{x}^{(j)} ; \myVec{s}^{(j)} \Big).
	\vspace{-0.1cm}
	\end{equation} 
	 
	When $\mySet{S}$ is a continuous set, which is the case in the ultrasound beamforming application discussed in Section~\ref{sec:AppBF}, then the system task is estimation over a continuous domain. Here, the output layer of the digital network is comprised of $\lenS$ nodes, each estimating a single entry of $\myS$. In such setups, the network output $\NetMap(\myX)$ takes values in $\mySet{S}^\lenS$, and is used as the estimate of $\myS$. A common loss function for such tasks is the empirical \ac{mse}, given by 
	\vspace{-0.1cm}
	\begin{equation}
	\label{eqn:LossFuncMSE}
	\mathcal{L}_{\Acqparams}(\Hparams)  =\frac{1}{\Ntraining}\sum_{j=1}^{\Ntraining} \Big\|  \NetMap\Big(\myVec{x}^{(j)}  \Big) - \myVec{s}^{(j)} \Big\|^2.
	\vspace{-0.1cm}
	\end{equation} 

	Once the system is trained, the learned parameters are used to configure the task-based acquisition system detailed in Subsection \ref{subsec:Model_Acquisition}, which operates on \ac{ct} signals. The matrix representation of the linear analog network is used to set the filter $\myH(t,\tau)$. The adaptation of the \ac{adc} mapping,  trained along with the overall system assuming a fixed number of samples $\SmpSize$ and quantization resolution $\Qres$, is detailed in the sequel, while a method for optimizing the acquisition hyperparameters $\Acqparams$ is discussed in Section \ref{sec:Meta}.

	\vspace{-0.2cm}
	\subsection{Learned Analog-to-Digital Conversion}
	\label{subsec:Layers}
	\vspace{-0.1cm}
	The analog and digital processing parts of the hybrid task-based acquisition system are learned as conventional \ac{dnn} models. However, the \ac{adc}  
	mapping, determined by the sampling and quantization rules whose adjustable parameters  are the sampling times $\{t_j\}$ in  $\Samp(\cdot)$ and the decision regions $\{a_i, b_i\}$ in $\Quan(\cdot)$,  
	cannot be represented using standard layers or activation functions. In particular, both the sampling mapping \eqref{eqn:Sampler} and the quantization function \eqref{eqn:quantizer} are non-differentiable or have a zero-valued gradient with respect to their input and/or the adjustable parameters. Consequently, one cannot use straight-forward application of backpropagation with \ac{sgd}-based optimization to train the system end-to-end.
	
	Following  \cite{shlezinger2019deep}, which considered quantization without sampling, we adopt a {\em soft-to-hard} approach. This approach approximates non-differentiable mappings {\em during training} by smooth functions that faithfully capture their operation. Recall that Kronecker delta functions, from which the sampling mapping in \eqref{eqn:Sampler} is comprised, can be obtained as the limit of a sequence of Gaussian functions with decaying variance. Substituting this  into \eqref{eqn:Sampler} while replacing  integration  with a summation over the discretized grid yields the following relaxation: 
	\begin{equation}
	\label{eqn:AppSampler}
	\left( \SampTr(\alpha(t))\right)_j = \sum_{i=0}^{\NyqSize-1} \alpha(i\NyqInt) \exp\left( \frac{\left( i\NyqInt - t_j\right)^2 }{\sigma^2_i}\right), 
	\end{equation}
	where the parameters $\{\sigma^2_i\}$ control the resemblance of  $\SampTr(\cdot)$ to the non-differentiable sampling function $\Samp(\cdot)$. Approximating $\Samp(\cdot)$ with \eqref{eqn:AppSampler} during training allows the system to learn the sampling  time instances $\{t_j\}$ along with the analog and digital networks. As samples are taken from the   grid $\{l\NyqInt\}$, the learned sampling times are projected onto this grid after training, i.e., $t_j$ is replaced with its nearest grid point, while each point can only be assigned to a single entry of $\{t_j\}$. 
	
\begin{figure}
		\centering
		\includefig{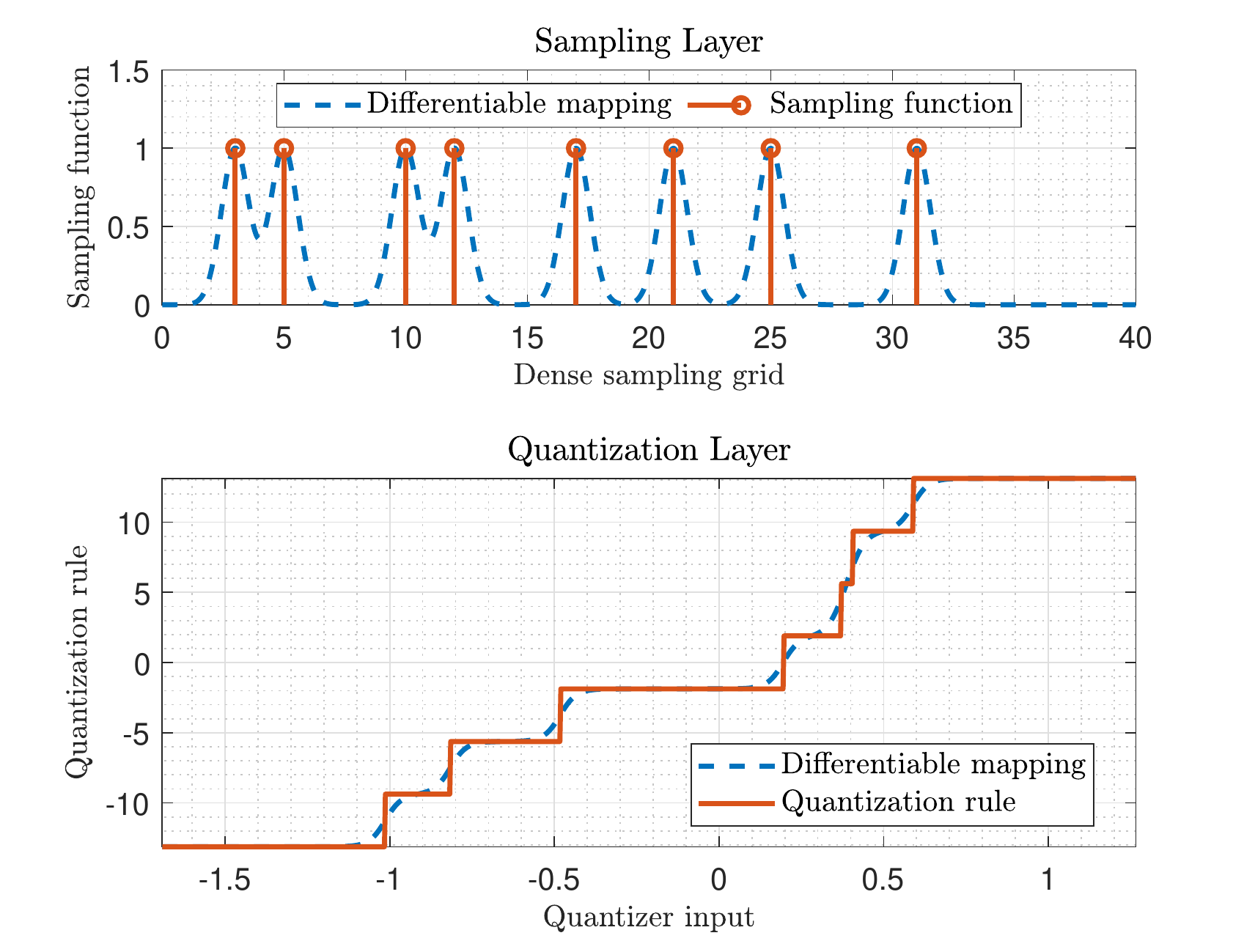} 
		\caption{Learned sampling and quantization rules illustration.}
		\label{fig:SoftToHard}
	\end{figure}

	Similarly, as suggested in \cite{shlezinger2019deep}, the quantizer \eqref{eqn:quantizer} is also approximated with a differentiable function. Since sign functions can be approached almost everywhere on $\mySet{R}$ by a sequence of hyperbolic tangents, $\Quan(\cdot)$ is approximated during training as 
	\begin{equation}
	\label{eqn:AppQuantizer}
	\QuanTr(\alpha) = a_0 + \sum_{i=1}^{\Qres-1}a_i {\rm tanh}\left( c_i\cdot\alpha - b_i\right), 
	\end{equation}
	where $\{c_i\}$ is a set of real-valued parameters. As $c_i$ increases, its corresponding ${\rm tanh}$ function approaches a sign mapping as in  \eqref{eqn:quantizer}. Using  \eqref{eqn:AppSampler}-\eqref{eqn:AppQuantizer}, the system can tune its parameters by backpropagating the gradient through the \ac{adc} mapping during training, while learning non-uniform quantization mappings by tuning  $\{a_i, b_i\}$. Once training is concluded, the learned set $\big\{\frac{b_i}{c_i}\big\}$ is used to determine the borders of the decision regions of the true (non-differentiable) quantizer. 
	Fig.~\ref{fig:SoftToHard} illustrates how  $\SampTr(\cdot)$ and $\QuanTr(\cdot)$ are converted into  sampling and quantization rules. 
	
	In  \eqref{eqn:AppSampler} and \eqref{eqn:AppQuantizer}, the parameters $\{\sigma_i^2\}$ and $\{c_i\}$, respectively,  balance  the smoothness of the mapping and the accuracy in representing the non-differentiable function from which it originates. Consequently, they can be either fixed, or  modified during training using annealing optimization \cite{rose1992vector}, making the differentiable mapping gradually approach the actual non-differentiable function during  training. 
	
	\vspace{-0.2cm}
	\subsection{Discussion}
	\label{subsec:Discussion}
	\vspace{-0.1cm}
	The proposed deep task-based acquisition system jointly adapts its analog filter, sampling function, quantization rule, and digital processing based on training data. This is achieved by identifying smooth trainable approximations of the sampling and quantization mappings.  The number of \acp{adc} $\lenZ$, average sampling rate $\frac{\SmpSize}{\Interval}$, and the quantization resolution $\Qres$ are all assumed to be fixed here. Yet, the fact that they can be treated as hyperparameters of a \ac{dnn} motivates their setting via hyperparameter optimization, as we explore in Section~\ref{sec:Meta}. 

	The system model detailed in Section \ref{sec:Model} allows the analog filter to implement any linear processing. As a result, the learned matrix $\bar{\myMat{H}}$  can be any real-valued $\NyqSize \cdot \lenX \times \NyqSize \cdot \lenZ$ matrix. Nonetheless, in some cases, a specific family of filters, such as causal, \ac{lti}, memoryless, or phase-shifters may be preferred. Such restrictions can be incorporated into our model by imposing a specific structure on $\bar{\myMat{H}}$. For example, for \ac{lti} filters, $\bar{\myMat{H}}$ is block-Toeplitz as shown in Example~\ref{exm:BlockTop}, while for memoryless filters, $\bar{\myMat{H}}$ can be written as $\myI_{\NyqSize} \otimes \tilde{\myMat{H}}$ for some $\tilde{\myMat{H}} \in \mySet{R}^{\lenX \times \lenZ}$. Alternative forms of constrained analog processing, such as those induced by the inherent controllable combining of dynamic metasurface antennas \cite{shlezinger2019dynamic,wang2020dynamic,shlezinger2020dynamic} or by using phase shifter networks \cite{ioushua2019family,mendez2016hybrid}, result in different constraints on $\bar{\myMat{H}}$. 
	Here, we focus on generic linear analog mappings, and leave  these special cases to future investigation. 
	
	As detailed in Subsection \ref{subsec:Architecture},  our  system is designed to be trained  offline, and the learned parameters are configured in the acquisition system once training is concluded. Nonetheless, one  can also envision an adjustable acquisition hardware device, which is capable of learning its multivariate analog-to-digital conversion mapping online. Such a system can utilize configurable \acp{adc} combined with neuromorphic circuits \cite{mead1990neuromorphic} based on, e.g., memristors \cite{danial2018breaking}, for realizing the trainable analog network. However, for a deep task-based acquisition system to adjust its parameters online using conventional methods such as \ac{sgd}, the digital processor must have access to the vector $\myX$ during training, namely, it must process a high-resolution  version of the vector representation of its observed signals during the periods in which it has knowledge of the task $\myS$. This requirement can be satisfied by, e.g., utilizing additional dedicated high-resolution \acp{adc} which are employed only during the specific periods where the analog input can be used for training along with its  label. 

	\vspace{-0.2cm}
	\section{Acquisition Hyperparameters Optimization}
	\label{sec:Meta}
	\vspace{-0.1cm}
	
	So far, we have utilized \ac{ml} methods to jointly learn the analog filtering, \ac{adc} mappings, and  digital processing, in an end-to-end manner. This learning stage is carried out 
	while fixing some of the key parameters of analog-to-digital conversion: the number of \acp{adc} $\lenZ$, the number of samples taken $\SmpSize$, and the quantization resolution $\Qres$. The fact that these hyperparameters directly affect some important aspects of acquisition, such as power consumption and memory usage \cite{walden1999analog}, motivates their learning as part of the training procedure.  In this section we detail how the proposed framework of learned task-based acquisition can be extended to tune these key parameters using Bayesian meta-learning tools. To that aim, we first introduce some basics in Bayesian meta-learning in Subsection \ref{subsec:MetaPrelim}, after which we present a method for optimizing  the acquisition parameters in Subsection \ref{subsec:MetaOptim}.


	
	
	
	\vspace{-0.2cm}
	\subsection{Preliminaries in Meta-Learning}
	\label{subsec:MetaPrelim}
	\vspace{-0.1cm}	
	Meta-learning is a subfield of \ac{ml} which deals with optimizing hyperparameters. Unlike conventional parameters, e.g., the weights of a neural network, that  are learned in the training process, hyperparameters are parameters of \ac{ml} algorithms that control the model class, e.g., the network architecture \cite{vinyals2016matching}, or the learning process, e.g., the learning rate \cite{maclaurin2015gradient} and the optimization rule \cite{wichrowska2017learned}.  Those can be either chosen from a discrete set or from a continuous range; the space of hyperparameters is referred to henceforth as the {\em search space}. 
	
	Some hyperparmeters, such as the initial weights used during training and the learning rate, can be optimized using gradient based methods, as done in model-agnostic meta-learning \cite{finn2017model}. However, computing the gradient of the loss with respect to architecture-related hyperparameters, such as the acquisition parameters of our deep task-based acquisition system, is often infeasible. For such settings where gradient-based methods cannot be applied, several methods have been proposed in the literature for hyperparmeter optimization. To formulate the different strategies, we use $\Acqparams \in \Theta$ to denote the set of hyperparameters of the learning algorithm $\mySet{A}$, where $\Theta$ is the search space. We also let $f_{\mySet{A}}(\Acqparams)$ be the evaluation function of these hyperparmeters. In the context of meta-learning the architecture of a neural network, computing this function involves training a network to obtain the empirical loss of the learning algorithm with hyperparmaters  $\Acqparams$. 
	Thus, computing  $f_{\mySet{A}}(\Acqparams)$ tends to be costly and time-consuming to evaluate.  Since for our task-based acquisition system, the search space $\Theta$ can be very large, while evaluating $f_{\mySet{A}}(\Acqparams)$ involves re-training a \ac{dnn} and is thus costly to compute, we adopt the Bayesian optimization approach for meta-learning~\cite{frazier2018tutorial}. 

	Bayesian meta-learning  involves a controllable amount of evaluations of $f_{\mySet{A}}(\cdot)$, and is based on assuming a prior distribution on it.   Bayesian  meta-learning models $f_{\mySet{A}}(\Acqparams)$ as a Gaussian process over $\Theta$ with postulated mean  $\mu_0(\Acqparams) = \E[f_{\mySet{A}}(\Acqparams)]$ and autocovariance function $\Sigma_0(\Acqparams, \Acqparams') = {\rm cov}\big(f_{\mySet{A}}(\Acqparams), f_{\mySet{A}}(\Acqparams') \big)$. The method sequentially samples the evaluation function $f_{\mySet{A}}(\cdot)$, iteratively refining the selected $\Acqparams$ assuming an underlying Gaussian model. Given the samples of $f_{\mySet{A}}(\cdot)$ measured at the $i$ vectors $\Acqparams_1, \ldots, \Acqparams_i$, the next sampling vector $\Acqparams_{i+1}$ is selected as the one maximizing the expected improvement \cite{frazier2018tutorial} 
	 	\begin{equation}
	 	\label{eqn:EI1} 
	    {\rm EI}_i(\Acqparams) \triangleq \mathbb{E}\Big[\big(f_\mathcal{A}(\Acqparams)-f_\mathcal{A}(\Acqparams_i^*)\big)^+|\{f_{\mySet{A}}(\Acqparams_j)\}_{j\leq i} \Big],
	\end{equation} 
	where $\Acqparams_i^*\triangleq \arg\max_{j\leq i} f_\mathcal{A}(\Acqparams_j)$ is the current best observation, and $a^+\triangleq \max(a,0)$. 
	
	Under the assumption that $f_\mathcal{A}(\cdot)$ is a Gaussian process, the  distribution of $f_\mathcal{A}(\Acqparams)$ conditioned on $\{f_{\mySet{A}}(\Acqparams_j)\}_{j\leq i}$ is also Gaussian with mean value $\mu_i(\Acqparams)$ and standard deviation $\sigma_i(\Acqparams)$, obtained from $\mu_0(\cdot)$ and $\Sigma_0(\cdot, \cdot)$ via
	\cite[Eq. (3)]{frazier2018tutorial}: 
	\begin{align*}
	    \mu_i(\Acqparams) &=  \Sigma_0(\Acqparams,\Acqparams_{1:i})\Sigma_0^{-1}(\Acqparams_{1:i},\Acqparams_{1:i})\notag \\
	    &\qquad \qquad \times \big(f_\mathcal{A}(\Acqparams_{1:i})-\mu_0(\Acqparams_{1:i})\big) + \mu_0(\Acqparams), \\
	   \sigma_n^2(\Acqparams) &= \Sigma_0(\Acqparams,\Acqparams)-\Sigma_0(\Acqparams,\Acqparams_{1:i})\Sigma_0^{-1}(\Acqparams_{1:i},\Acqparams_{1:i})\Sigma_0(\Acqparams_{1:i},\Acqparams). 
	\end{align*} 
	Here,  $f_\mathcal{A}(\Acqparams_{1:i})$ denotes the $i\times 1$ vector whose $i$th entry is $f_{\mathcal{A}}(\Acqparams_i) $;
	$\Sigma_0(\Acqparams_{1:i},\Acqparams_{1:i})$ is an $i \times i$ matrix whose $(m,j)$th entry is $\Sigma_0(\Acqparams_{m},\Acqparams_{j})$;   $\Sigma_0(\Acqparams,\Acqparams_{1:i})$ is a $1 \times i$ vector whose $j$th entry  is $\Sigma_0(\Acqparams,\Acqparams_{j})$; and $\mu_0(\Acqparams_{1:i})$ is an $i\times 1$ vector whose $i$th entry is $\mu_0(\Acqparams_{i})$.   
 By letting  $F_{G}(\cdot)$ and $p_{G}(\cdot)$ are the cumulative distribution function and the probability density function of the standard normal distribution, respectively, the expected improvement is computed via \cite[Eq. (8)]{frazier2018tutorial}
	\begin{equation}
    \label{eqn:ExpectedImprovment}
	    {\rm EI}_i(\Acqparams) \!=\! 
	    (\mu_n(\Acqparams) \!-\! f_\mathcal{A}(\Acqparams_i^*))F_{G}(Z_i) \!+\! \sigma_i(\Acqparams)p_{G}(Z_i),
\end{equation}
where
\begin{equation}
	    Z_i = \begin{cases}
	    \frac{(\mu_i(\Acqparams) - f_\mathcal{A}(\Acqparams_n^*).)}{\sigma_i(\Acqparams)}
          &\mbox{if } \sigma_i(\Acqparams) > 0, \\  
        -\infty &\mbox{if } \sigma_i(\Acqparams) = 0.
    \end{cases}
\end{equation}

The resulting sequential hyperparameter optimization is summarized as Algorithm \ref{alg:Algo1}. The term  $f_\mathcal{A}(\Acqparams_i^*)$ is often replaced with $f_\mathcal{A}(\Acqparams_i^*)+\zeta$ for some $\zeta > 0$ to improve exploration and reduce the probability of yielding a local optima \cite{brochu2010tutorial}.

    \begin{algorithm}  
		\caption{ Bayesian hyperparameters optimization}
		\label{alg:Algo1}
		\KwData{Randomly sample hyperparameter configuration \(\Acqparams_1\), set \(\Acqparams^* = \Acqparams_1\), calculate \(f_\mathcal{A}(\Acqparams_1)\)}.
		\For{$i=1,2,\ldots,i_{\max}-1$}{
		Compute \(\Acqparams_{i+1} = \argmax_{\Acqparams} {\rm EI}_i(\Acqparams)\) via \eqref{eqn:ExpectedImprovment}\\
	     Calculate \(f_\mathcal{A}(\Acqparams_{i+1})\) \\
	     \If{\(f_\mathcal{A}(\Acqparams^*) < f_\mathcal{A}(\Acqparams_{i+1})\)}{
    Update \(\Acqparams^* = \Acqparams_{i+1}\) } 
        }
    \KwOut{ Hyperparameter configuration \(\Acqparams^*\).}
	\end{algorithm}

    
	
	\vspace{-0.2cm}
	\subsection{Acquisition Parameters Optimization }
	\label{subsec:MetaOptim}
	\vspace{-0.1cm}
	Our proposed deep task-based \ac{adc} system is modeled as a \ac{dnn} such that its analog filter,  acquisition  mappings,  and  digital  processing can be learned  end-to-end from data. The analog-to-digital conversion configuration is dictated by  the number of \acp{adc} $\lenZ$, the number of samples taken in each interval $\SmpSize$, and the quantization resolution $\Qres$. The triplet $(\lenZ, \SmpSize, \Qres)$, which dictates  the number of bits used in  acquisition $\lenZ \cdot \SmpSize \cdot \lceil\log_2 \Qres\rceil$, affects the architecture of the \ac{dnn}, and is thus treated as the hyperparameters of the model. Consequently, in order to optimize the model under a given bit budget $\BitBudget$, as requested in the problem formulation in Subsection~\ref{subsec:Model_Problem}, we utilize meta-learning via Bayesian optimization. 
	By letting $\Theta_{\BitBudget}$ be the set of triplets of positive integers $(\lenZ, \SmpSize, \Qres)$ such that $\lenZ \cdot \SmpSize \cdot \lceil\log_2 \Qres\rceil \leq \BitBudget$, meta-learning is expressed as the following optimization problem: 
	\begin{equation}
    	 \mathop{\arg\min}\limits_{(\lenZ, \SmpSize, \Qres) \in  \Theta_{\BitBudget}} f_\mathcal{A}(\lenZ,\SmpSize,\Qres) \label{eqn:MetaOpt}
	\end{equation}
	
	The  objective in \eqref{eqn:MetaOpt} is determined by the system task and the available data set. For example, $f_\mathcal{A}(\lenZ,\SmpSize,\Qres)$ can  represent the the training loss as in \eqref{eqn:LossFuncCE}, e.g., 
		\begin{equation}
	\label{eqn:MetaObj}
	    f_\mathcal{A}(\Acqparams) = -\min_{\Hparams}\mySet{L}_{\Acqparams}(\Hparams),
	\end{equation}
	or alternatively, the validation error of a deep task-based acquisition system as in Fig. \ref{fig:SystemModel} with hyperparameters $(\lenZ,\SmpSize,\Qres)$  after trained using a given data set. 
	Further, one can boost configurations with reduced number of overall bits by including in the formulation of the objective a regularization term which accounts for the number of bits acquired, as we do in the numerical study in Subsection~\ref{subsec:MIMOSimsMeta}. 
	
	The objective in \eqref{eqn:MetaOpt} satisfies the following conditions: $1)$ Computing  $f_\mathcal{A}(\cdot)$ requires training the network anew, and thus involves a computationally expensive computation; $2)$ the cardinality of the search space $\Theta_{\BitBudget}$ grows with the size of the sampling grid $\NyqSize$, and can thus be large. These conditions imply that naive search techniques may be computationally infeasible, hence,  we utilize Bayesian meta-learning in Algorithm \ref{alg:Algo1} for tuning the \ac{adc} configuration. 	The resulting overall learning procedure of the deep task-based \ac{adc}, including the learning of the both the acquisition hyperparameters $\Acqparams$ and the \ac{dnn} weights $\Hparams$, is summarized as Algorithm~\ref{alg:Algo2}. 
	
	  \begin{algorithm}  
		\caption{ Deep task-based \ac{adc} learning}
		\label{alg:Algo2}
		\KwData{Define loss measure $\mySet{L}_{\Acqparams}(\cdot)$ and meta-learning objective 
		$f_\mathcal{A}(\Acqparams)$. }
		Obtain $\Acqparams$ via Algorithm~\ref{alg:Algo1}. \label{stp:MetaTrain}\\
		Set $\Hparams$ to minimize $\mySet{L}_{\Acqparams}(\cdot)$ via  \ac{dnn} training.  \label{stp:Train} \\
    \KwOut{ Deep task-based \ac{adc} system $\NetMap(\cdot)$.}
	\end{algorithm}
	
	This application of Algorithm~\ref{alg:Algo2} can be further facilitated by recasting the multiplicative formulation of the search space $\Theta_{\BitBudget}$ into an additive one by writing it as
		\begin{equation}
	\label{eqn:BitBudgetConstraint}
	  \log_2\lenZ + \log_2\SmpSize + \log_2\lceil\log_2\Qres\rceil \leq \log_2\BitBudget. 
	\end{equation}
	Expressing $\Theta_{\BitBudget}$ via \eqref{eqn:BitBudgetConstraint} enables the application of existing Bayesian meta-learning toolboxes, such as BoTorch \cite{Balandat2019botorch} and  Ax \cite{ax}; the latter is used in  our numerical evaluations.  Algorithm~\ref{alg:Algo2} can be further simplified by noting that the Bayesian optimization procedure involves training the \ac{dnn} with the optimized acquisition hyperparameters, and thus one can extract the network weights $\Hparams$ from Algorithm~\ref{alg:Algo1} and avoid re-training in Step~\ref{stp:Train}, as we do in our numerical study in Subsection~\ref{subsec:MIMOSimsMeta}. In fact, such an approach is expected to yield improved performance as the networks weights are selected from a set of $i_{\max}$ independent training procedures. Nonetheless, as one may prefer to utilize different data sets or different optimization configuration (e.g., step size, number of epochs) in learning the \ac{dnn} weights compared to those used when evaluating the meta-learning objective $f_\mathcal{A}(\cdot)$, we include a dedicated separate weights training step in the overall learning procedure in Algorithm~\ref{alg:Algo2}.

	\vspace{-0.2cm}
	\section{Case Study: Synthetic Model }
	\label{sec:AppMIMO}
	\vspace{-0.1cm}
	We next apply the deep task-based acquisition system for detection in a synthetic linear model. The aim of this study is to demonstrate the ability of the proposed mechanism to jointly train the component of the hybrid analog/digital system, as well as to evaluate the hyperparameter optimization mechanism proposed in Section~\ref{sec:Meta}. Therefore, in this section we consider a relatively simple synthetic model for which we are able to, e.g., compute the model-based \ac{map} rule as a benchmark. The application of the proposed deep task-based acquisition framework in a non-synthetic setup is detailed in next case study in Section~\ref{sec:AppBF}.
	
	Here, we first describe the task and the signal model and the experimental setup\footnote{The source code used in our numerical studies is available online on \url{https://github.com/arielamar123/ADC-Learning-hyperopt}.} in Subsection \ref{subsec:Experimental}. Then, we evaluate the  deep task-based acquisition system  for such setups with fixed acquisition hyperparameters in Subsection \ref{subsec:MIMOSims} and with optimized hyperparameters in Subsection \ref{subsec:MIMOSimsMeta}. 

	\vspace{-0.2cm}
	\subsection{Experimental Setup}
	\label{subsec:Experimental}
	\vspace{-0.1cm}
	
	We consider the detection of a vector of binary-valued symbols $\myVec{s}$ whose entries take value in a discrete set  $\mySet{S} = \{-1, 1\}$. The \ac{ct} $\lenX \times 1$ signal $\myX(t)$ observed  at time instance $t \in [0,\Interval)$ is related to the task vector $\myVec{s}$ via the following linear model:
	\vspace{-0.1cm}
	\begin{equation}
	\label{eqn:LinearChannel}
	\myX(t) = \myMat{G}(t)\myS + \myVec{w}(t).
	\vspace{-0.1cm}
	\end{equation}
	Here, $\myMat{G}(t) \in \mySet{R}^{\lenX \times \lenS}$ is time-varying measurement matrix and $\myVec{w}(t)$ is the noise vector, comprised of independent zero-mean  Gaussian entries with variance $\sigma_w^2(t) >0$. To obtain the signal model as in \eqref{eqn:XModel2}, we approximate the signal the \ac{ct} using dense sampling, such that $\tilde{\myX}_j = \myX(j\cdot \Interval/\NyqSize)$.
	
	The task is thus given by the  recovery of $\myS$ from the observed $\myX(t)$, and can thus be treated as acquisition for a classification task. Note that in the absence of noise, $\myS$ can be often accurately recovered from a single sample of $\myX(t)$, and thus the gain in processing multiple samples is in reducing the effect of noise quantization distortion. 
We set $\lenX = 6$, $\lenS = 4$,  and the signal duration is $\Interval = 1$ $\mu{\rm Sec}$. The noise  in \eqref{eqn:LinearChannel} satisfies $\sigma_w^2(t) \equiv 1$, while the measurement matrix $\myMat{G}(t)$ represents spatial exponential decay with temporal variations, and its entries are 
	\vspace{-0.1cm}
	\begin{equation}
	\label{eqn:ChannelMat}
	\left( \myMat{G}(t)\right)_{i,j} = \sqrt{\rho}(1+0.5\cos(2\pi f_0t))e^{-|i-j|},
	\vspace{-0.1cm}
	\end{equation}
	where $\rho > 0$ is referred to as the \ac{snr} and $f_0 = 10^3$ Hz. Note that the fact that the measurement matrix  \eqref{eqn:ChannelMat} varies within the symbol duration motivates the usage of non-uniform sampling. 
	
	In our experimental study we implement the following architecture for the deep task-based analog-to-digital conversion system:  
 The analog network is an  $\lenX\cdot\NyqSize \times \lenZ \cdot \NyqSize$ fully-connected layer, and  the digital \ac{dnn}, comprised of a $\lenZ \cdot \SmpSize \times 32$ layer,  a ReLU activation, a $32 \times 16$ layer, and a softmax output layer.
 The network is trained 	to minimize \eqref{eqn:LossFuncCE}
	over $\Ntraining = 10^4$ samples using the ADAM optimizer \cite{kingma2014adam} with learning rate of $0.01$.

	\vspace{-0.2cm}
	\subsection{Fixed Hyperparameters Experiments}
	\label{subsec:MIMOSims}
\begin{figure}
		\centering
		\includefig{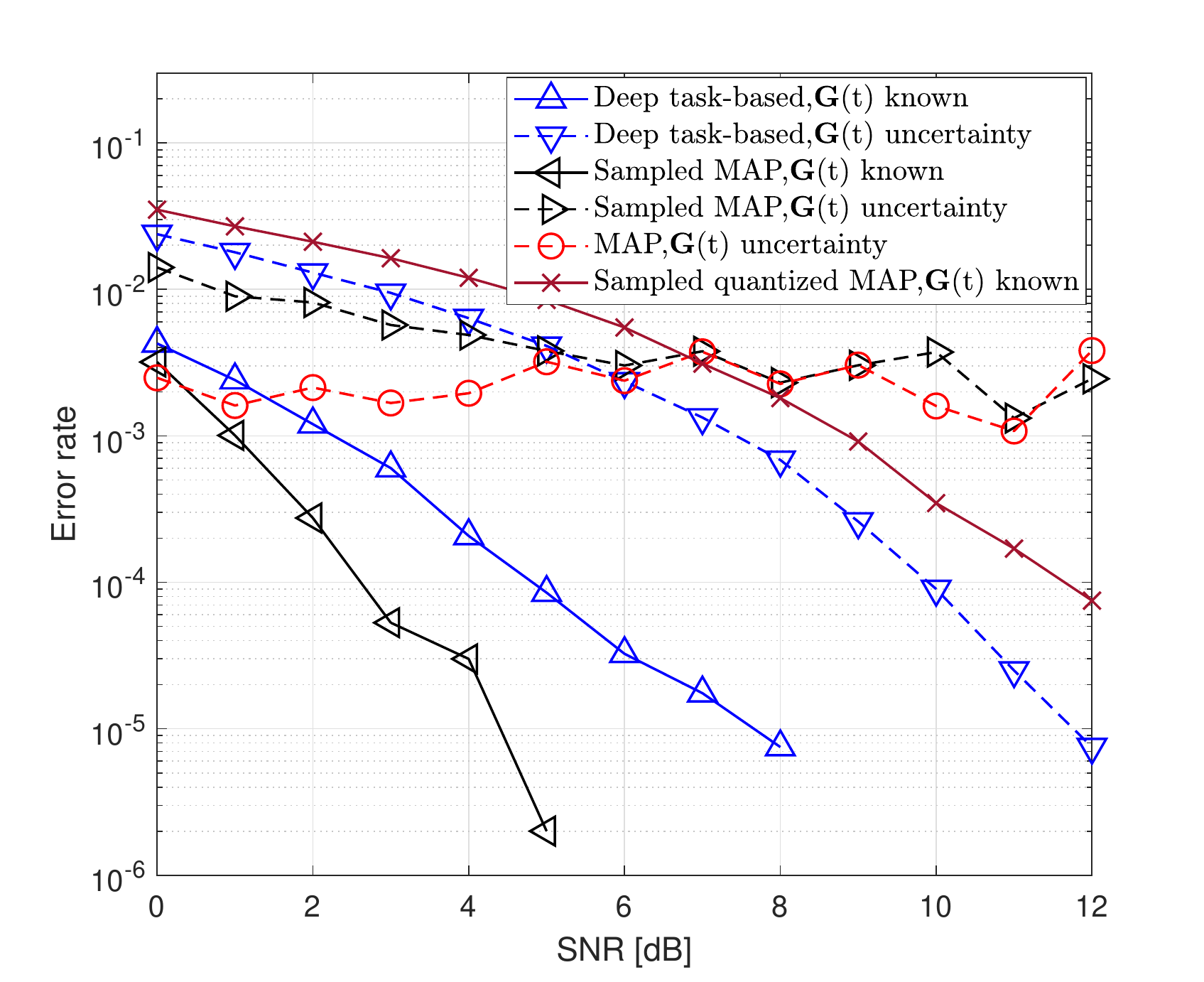} 
		\vspace{-0.2cm}
		\caption{Error rate versus \ac{snr}.}
		\label{fig:BERvsSNR_R2}
	\end{figure}

	\vspace{-0.1cm} 
	We begin by evaluating the deep task-based acquisition system with fixed acquisition hyperparameters. The simulated acquisition system uses $\lenZ = 4$ \acp{adc}, while selecting $\SmpSize = 4$ samples out of a  grid of $\NyqSize=20$ time instances, and quantizing each sample using up to $\log_2 \Qres = 3$ bits.
	We compare the error rate of our deep acquisition system to the following model-based detectors: The \ac{map} rule for recovering $\myS$   from a uniformly sampled version of $\myX$  with sampling rate $\SmpSize / \Interval$ referred to as {\em sampled \ac{map}}, namely, the minimal achievable error rate when using the same number of samples as our deep task-based system without quantization constraints; and the \ac{map} rule for recovering $\myS$ from $\myX$ from a uniformly sampled and quantized version of $\myX$ without analog processing, referred to as {\em sampled quantized \ac{map}}. The resulting error rates, averaged over $10^5$ Monte Carlo simulations, versus \ac{snr} are depicted in Fig.~\ref{fig:BERvsSNR_R2}.
	
	We note that while the data-driven acquisition system is ignorant of the statistical model relating $\myS$ and $\myX(t)$ and learns its mapping from training samples corresponding to this model, the  \ac{map} receivers require accurate knowledge of the underlying model. In particular, they rely on the fact that the underlying signal model is given by \eqref{eqn:LinearChannel}, and require knowledge of $\myMat{G}(t)$. As accurate knowledge of the signal model may not be available in some scenarios, we also depict in Fig. \ref{fig:BERvsSNR_R2} the error rate obtained by the sampled \ac{map} receiver as well as the \ac{map} rule for recovering $\myS$ from $\myX$ without sampling and quantization constraints, when these receivers have access to a noisy version of $\myMat{G}(t)$, in which each entry  is corrupted by additive i.i.d. Gaussian noise whose variance is $30\%$ of its magnitude. This scenario is referred to as $\myMat{G}(t)$ uncertainty. To evaluate our  acquisition system under $\myMat{G}(t)$ uncertainty, we compute its achievable error rate when trained using samples taken from the same inaccurate noisy signal model. 
	
	Observing Fig. \ref{fig:BERvsSNR_R2}, we note that for accurate training, our deep task-based acquisition system achieves comparable performance to the sampled \ac{map} which operates without quantization constraints. Furthermore, our data-driven system notably outperforms the quantized \ac{map} rule, which utilizes uniform quantizers of lesser resolution, as it does not reduce the dimensionality in analog  and must thus assign less bits for each \ac{adc}.
	In the presence of model uncertainty, the performance of our proposed system is degraded by approximately $5$~dB in \ac{snr} compared to accurate training, yet it is still capable of achieving error rate  below $10^{-4}$ for \acp{snr} above $10$~dB. The model-based \ac{map} rule operating without quantization constraints, whether processing a uniformly sampled input or even on the densely discretized $\myX$, reaches an error floor of above $10^{-3}$. These results demonstrate the ability of the proposed deep task-based acquisition framework in jointly optimizing the analog and digital mappings along with the \ac{adc} rule in a manner which allows to accurately carry out the desired task. 
	
    \begin{figure*}
        \centering     
        \subfigure[Synthetic signal model, $k=4$ and $n=6$ ]{\label{fig:a}\includegraphics[width=80mm]{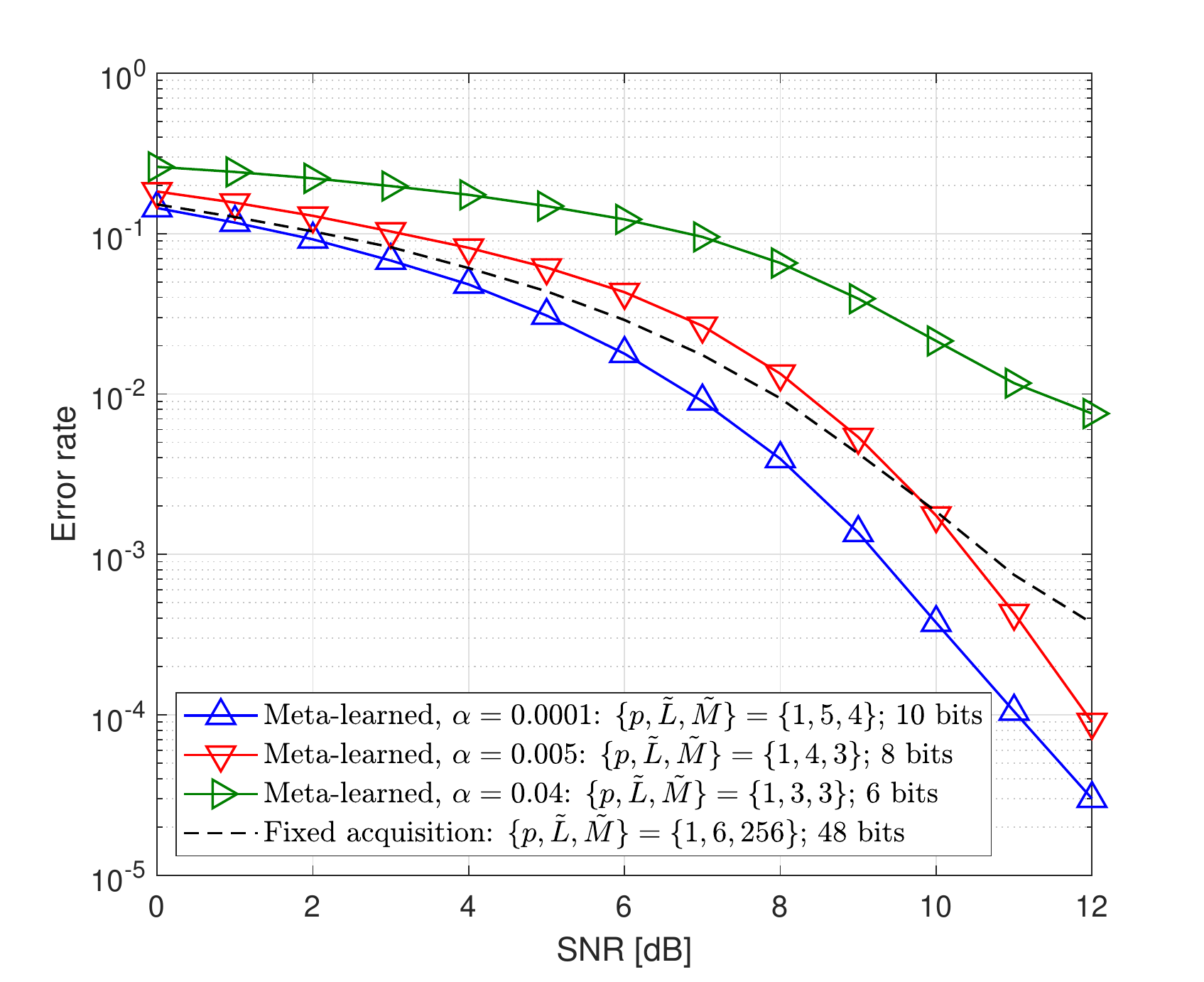}} 
        \quad
        \subfigure[Synthetic signal model, $k=8$ and $n=16$]{\label{fig:b}\includegraphics[width=80mm]{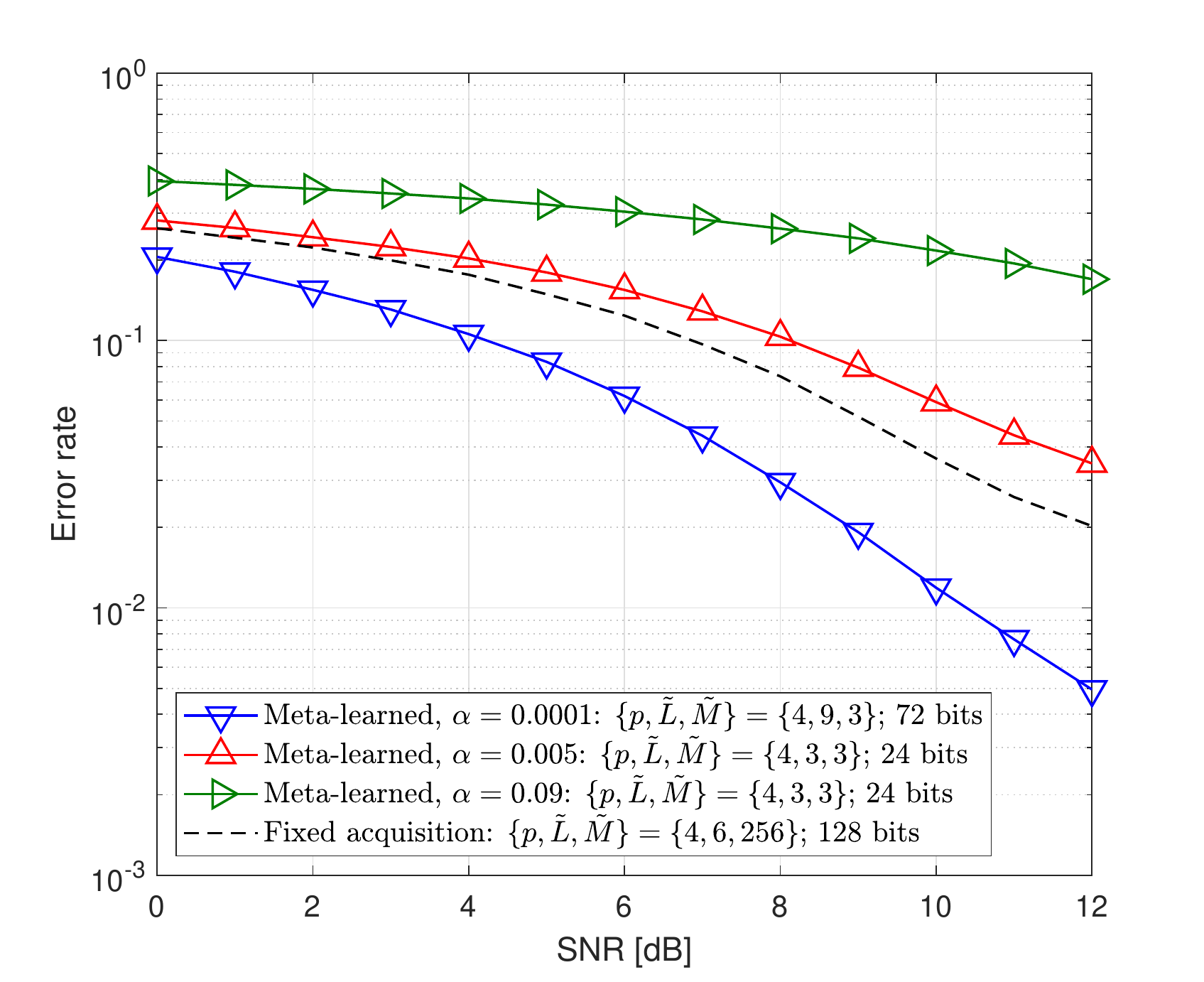}}
        \caption{Error rate versus SNR for deep task-based acquisition with meta-learned acquisition parameters versus fixed parameters}
        \label{fig:BER_vs_SNR_hyperparameters}
    \end{figure*}
    
    \begin{figure}
    	\centering
    	\includegraphics[width=\linewidth]{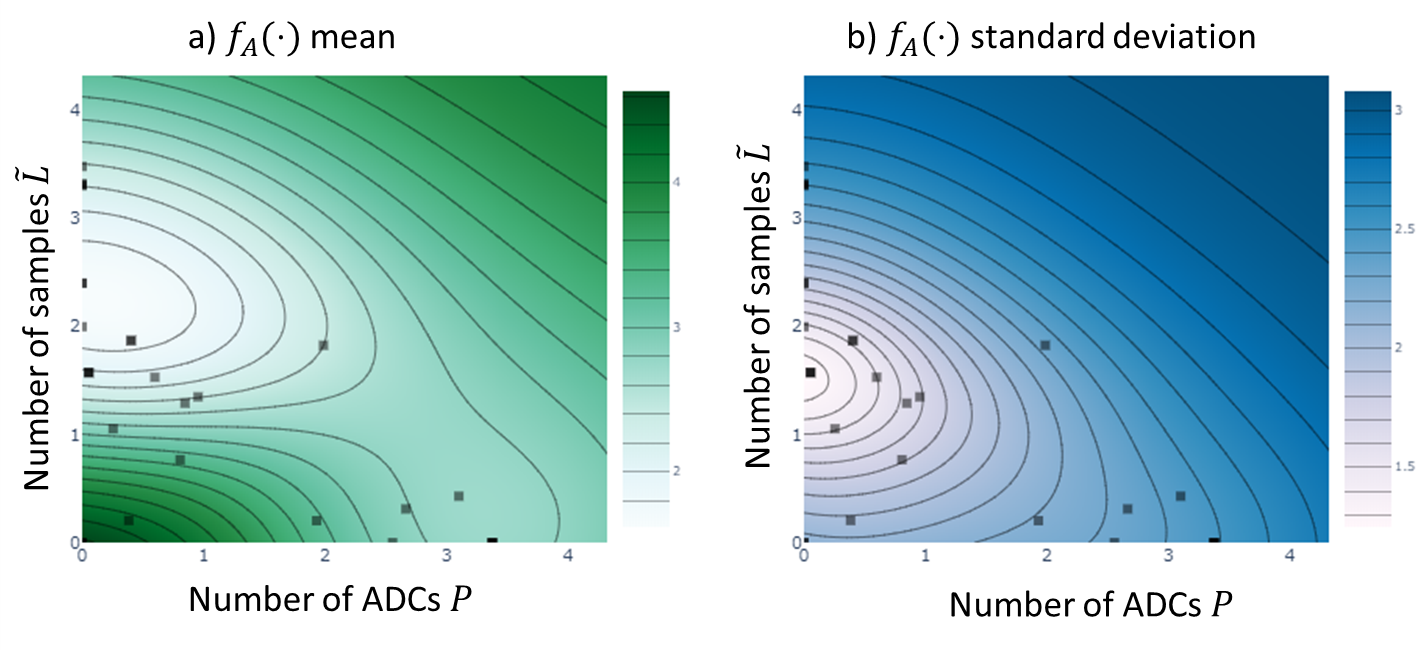} 
    	\vspace{-0.8cm}
    	\caption{Contour plots of the  $(a)$ mean  and $(b)$ standard deviation  of \(f_\mathcal{A}\) as function of number of \acp{adc}, and number of samples when quantization resolution is fixed \(\Qres = 4\). Squares marked values of \(f_\mathcal{A}\) queried in hyperparameter optimization}
    	\label{fig:P_vs_L_tilde}
    \end{figure}	
	
	\vspace{-0.2cm}
	\subsection{Meta-Learned Acquisition Hyperparameters}
	\label{subsec:MIMOSimsMeta}
	\vspace{-0.1cm}
	Next, we numerically evaluate the meta-learning procedure detailed in Section \ref{sec:Meta} for optimizing the acquisition hyperparameters. We do so by applying Algorithm~\ref{alg:Algo1} to optimize the parameters of the acquisition system, such that the overall number of bits utilized is minimized without degrading the overall performance. The network weights are selected as the ones trained along with the selected hyperparameters in the Bayesian optimization procedure, as discussed in Subsection~\ref{subsec:MetaOptim}.
	
	Here, we fix the maximal overall bit budget to  \(\BitBudget = 20\) bits. Since hyperparameters optimization does not involve computing the gradient of its objective $f_\mathcal{A}(\Acqparams)$, we  use the error rate objective, being the desired performance measure, rather than the cross entropy loss \eqref{eqn:LossFuncCE} which is used when training the weights. In particular, we design the Bayesian optimization procedure to both tune \(\lenZ,\SmpSize,\Qres\) to get low error rate as possible , and use lowest possible number of bits under the constraint of our bit budget. The objective is thus set to
	\begin{align}
	  \Acqparams^* \triangleq (\lenZ^*, \SmpSize^*, \Qres^*) =  &\mathop{\arg\min}\limits_{\lenZ \cdot \SmpSize \cdot \lceil\log_2(\Qres)\rceil \leq \BitBudget}  \alpha\cdot(\lenZ \cdot \SmpSize \cdot \lceil\log_2(\Qres)\rceil)  \notag \\ 
	  &+ \sum_{\rho \in \mathcal{P}} {\rm ER}_\rho(\NetMap),
	  \label{eqn:objective}
	\end{align}
	where
	 $\mathcal{P}$ is the set of SNR values for each channel we are testing; \({\rm ER}_{\rho}(\NetMap)\) is the error rate achieved using an an acquisition system with hyperparameters $\Acqparams$ that was trained with  channel with SNR \(\rho\); and \(\alpha\) balances the contribution of two measures one is interested in minimizing: the number of bits the acquisition system is using and the model performance respectively.

In the setting with fixed hyperparameters detailed in Subsection~\ref{subsec:MIMOSims}, the acquisition system uses $\lenZ = 4$ \acp{adc}, $\SmpSize = 4$ samples out of a grid of $\NyqSize=20$ time instances, and samples are quantized using $\log_2 \Qres = 3$ bits. Thus, an overall of \(4\cdot4\cdot3 = 48\) bits are used for acquisition.
As observed in Fig. \ref{fig:a}, the proposed Bayesian meta-learning scheme allows to achieve error rates results within a minor gap of that of the original configuration while using 62.5\% less bits.
In Fig. \ref{fig:b}, we illustrated that for larger scale settings, with $k=8$ and $n=16$, hyperparameter optimization via Algorithm~\ref{alg:Algo2} manages to reduce the number of bits by 75\% with only a small loss in the error rate, while even allowing to achieve improved accuracy when the number of bits is reduced by $45\%$, due to its inherent training of multiple systems and the selection of the most accurate one. These results indicate that the proper combination of Bayesian meta-learning with learning of the overall mapping via deep task-based acquisition allows to improve both performance and bit efficiency.
	
In Fig. \ref{fig:P_vs_L_tilde}	we depict a contour plot representing the objective \(f_\mathcal{A}(\theta)\) as a function of $\lenZ$ and $\NyqSize$ when fixing the quantization resolution to be $\Qres=4$. This plot shows the relations between number of \acp{adc} and number of samples taken when the quantization resolution is relatively small.
We can see from Fig. \ref{fig:P_vs_L_tilde} that Algorithm~\ref{alg:Algo2} is likely to prefer hyperparameter configurations taking small number of \acp{adc} with relatively large amount of samples. Another option shown in the plot is taking relatively large amount of \acp{adc} with low amount of samples, shown to be a local minimum of the objective function. However, when looking in the standard deviation contour plot we can see that a configuration like this will not be stable, because the standard deviation is high.

    \begin{figure*}[h]
		\centering
		\includegraphics[trim=1cm 1cm 1cm 1cm, width=18cm]{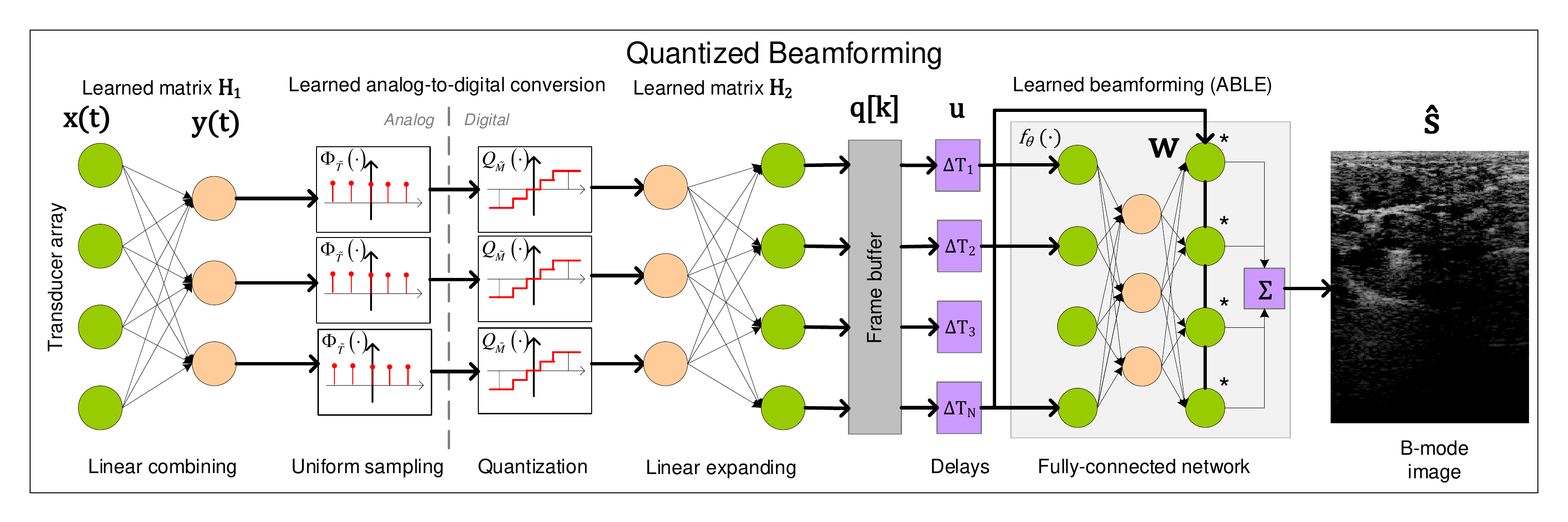}		
		\caption{Schematic overview of the task-based analog-to-digital conversion for ultrasound, jointly trained with adaptive beamforming by deep learning. The (analog) RF signals are consecutively compressed in $p$ channels, uniformly sampled, quantized, and expanded back into $n$ channels before being stored in a digital buffer. Trough delays, these RF lines can be focused to individual pixels, after which adaptive apodization (ABLE) is applied to yield a beamformed RF image.
		}
		\label{fig:US_system}
	\end{figure*}	
     	
	\vspace{-0.2cm}
    \section{Case Study: Ultrasound Beamforming} 
    \label{sec:AppBF}
    \vspace{-0.1cm}
    

    In this section we apply the deep task based acquisition framework to a real-world case study of ultrasound image reconstruction, which can be modeled as a regression problem. Here, we show that joint training of task-based \acp{adc} combined with deep learning based adaptive beamforming can achieve high quality imaging at low data rates, improving over competing approaches.

    \begin{figure*}
		\centering
		\includegraphics[trim=3cm 0cm 3cm 0cm,width=8.5cm]{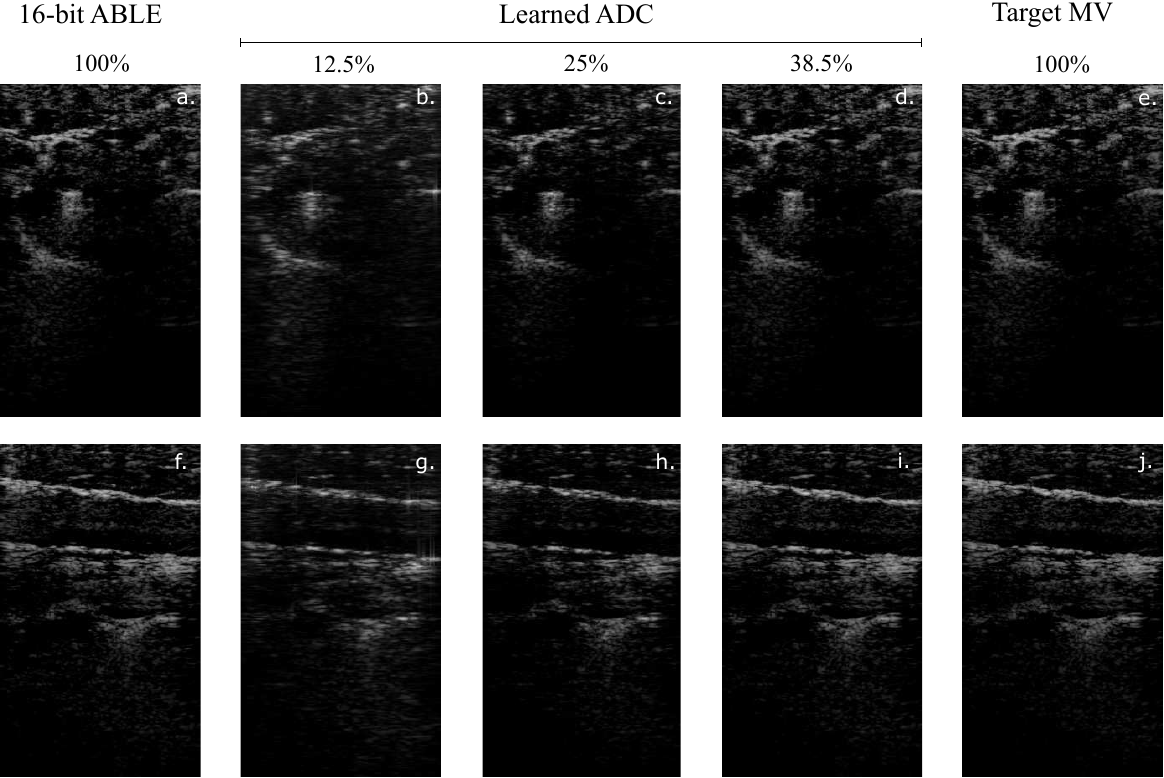}
		\caption{Image reconstructions, with a dynamic range of 60dB, for: a-e) Carotid artery cross-section, f-g) Carotid artery longitudinal cross-section. For each image, reconstructions are shown for: Uncompressed 16-bit ABLE (a \& f), the proposed task based acquisition with compression ratios of 12.5\% (b \& g), 25\% (c \& h), and 38.5\% (d \& i), and finally the 16-bit minimum variance beamformed training targets. (e \& j)}
		\label{fig:results_US}
	\end{figure*}
	
    
    Ultrasound imaging is based on the transmission and reflection of high frequency sound waves in tissue. These reflections, denoted $\{x_i(t)\}_{i=1}^{\lenX}$,  are recorded by an array of $n$ transducer elements, and are used to form a brightness mode (B-mode) image of the tissue by applying beamforming. Typical ultrasound devices use probes with $n=128$ or more transducer elements, each of which contributing a separate data-stream. Additionally, multiple consecutive recordings might be required for a single frame in order to achieve an SNR that leads to the desired image properties. The transfer of these recordings involves hardware that supports a large data bandwidth, which is expensive and not always feasible to implement. In recent developments such as 3D ultrasound ($n> 512$) or portable scanners (bandwidth constrained), these limitations are especially problematic. The need for compression of data early on in the signal chain motivates the use of our proposed deep task based signal acquisition in ultrasound imaging,  which is studied in this section, beginning the description of the experimental setup in Subsection~\ref{subsec:AppBFSetup}, and followed by the statement of the results in Subsection~\ref{subsec:AppBFResults}.
    
    \vspace{-0.1cm}
    \subsection{Experimental Setup}
    \label{subsec:AppBFSetup}
    \vspace{-0.1cm}
    
    \subsubsection{Data}
    We consider plane-wave (PW) imaging, in which a planar wavefront is transmitted, energizing the whole imaging medium with a single pulse. 
    For training we use 1000 \textit{in-vivo} recordings, acquired using a Verasonics Vantage system with the L11-4v linear probe. Additionally, 100 images are obtained for testing purposes.
    High quality target images were generated similar to \cite{luijten2020adaptive}, by computationally intensive minimum variance   beamforming. 
    
    \subsubsection{Filtering and quantization}
    Following the task-based acquisition model  depicted in Fig.~\ref{fig:SystemModel}, we consecutively filter, sample and quantize $\myVec{x}(t)$. During the forward pass, we follow Equations \eqref{eqn:AnalogFilt}, \eqref{eqn:Sampler} and hard quantization as in \eqref{eqn:quantizer}. The  input to the \acp{adc} is given by
    \begin{equation}
    \myVec{y}(t) = \myH_1 \myX(t) 
        \label{eqn:us_quantization}
    \end{equation}
    where 
    $\myH_1 \in \mathbb{R}^{p \times n}$ is a trainable matrix. Note that \eqref{eqn:us_quantization} specializes the generic formulation of \eqref{eqn:AnalogFilt} by restricting the time-varying analog filter $\myH(\cdot,\cdot)$ to represent time-invariant spatial combining, i.e., take the form $\myH(t,\tau) = \myH_1 \delta(\tau)$.  The filtering operation transforms the $n$ channel signals into a linear combination of $p$ channels, such that $p\leq n$, to further reduce datarates. 
%
    After quantization, the signals are expanded again into a set of $n$ signals through
    \begin{equation}
        \myVec{q}[i] = \myH_2 \Quan\Big( \myVec{y}(i T_L) \Big),
        \label{eqn:us_decode}
    \end{equation}
    where $\myH_2 \in \mathbb{R}^{n \times p}$ is another trainable  matrix that maps back to the original array geometry.

    To simplify notation, we summarize the sampling, quantization and filtering operations in \eqref{eqn:us_quantization} and \eqref{eqn:us_decode} as a single function $g_{\myVec{\phi}} (\cdot)$ such that
    \begin{equation}
        \myVec{q}[i] = g_{\myVec{\phi}} (\myX(iT_L)).
    \end{equation}
    The vector ${\myVec{\phi}}$ constitutes the trainable parameters, corresponding to the quantization levels and the filters $\myMat{H}_1, \myMat{H}_2$; the sampling instances in this experimental study represent fixed uniform sampling.
    
    \subsubsection{Beamforming}
    Next, the digital signals are focused towards each pixel position by applying delays, effectively transforming $\myVec{q}$ from the time-domain to the pixel-domain. For each pixel, this yields the $\lenX \times 1$ channel domain signal given by 
    \begin{align}
    \myVec{u}(\myVec{r}) &= 
    \left[u_1(\myVec{r}), u_2(\myVec{r}), \ldots, u_n(\myVec{r}) \right]\notag \\
    &= \left[\myVec{q}[\Delta_1(\myVec{r})], \myVec{q}[\Delta_2(\myVec{r})], \ldots, \myVec{q}[\Delta_\lenX(\myVec{r})] \right],
    \label{eqn:bf_delay}
    \end{align}
    where $\myVec{r}$ is a coordinate vector towards that pixel position, and $\Delta_i(\myVec{r})$ the corresponding time-delay.
    
    In conventional delay-and-sum beamforming (DAS), these delayed signals are weighed according to a window $\mathbf{w}(\myVec{r})$, favouring either contrast (Hanning window) or resolution (rectangular window), and subsequently summed. This operation is given by
    \begin{equation}
    \mathbf{\myVec{s}_{\text{DAS}}}(\myVec{r})=\myVec{w}^\textrm{T}(\myVec{r}) \mathbf{u}(\myVec{r}).
    \label{eqn:bf_das}
    \end{equation}
    where $\mathbf{s_\text{DAS}}$ denotes the beamformed signal output at every pixel index. Note here that, while $\myVec{r}$ can vary per pixel, it does not adapt to the received signals.

    To improve upon such a fixed apodization scheme, an adaptive method can be employed in the digital domain.
    Here, we consider the trainable Adaptive Beamforming by Deep Learning (ABLE) \cite{luijten2020adaptive}, a model-based deep learning framework which can learn to predict optimal channel apodizations based on time-delayed RF data.
    ABLE can be written as an apodization function $f_{\myVec{\theta}}(\cdot)$, which depends on a set of trainable parameters $\myVec{\theta}$, and maps an input signal to a content-adaptive apodization pattern. The ABLE model combined with the mapping matrix $\myMat{H}_2$ constitute the digital processing of the generic task-based acquisition system of Fig.~\ref{fig:GenSetup2}, as illustrated for the considered ultrasound beamforming setup in Fig.~\ref{fig:US_system}.

    To conclude, our beamformed output signal is given by 
    \begin{equation}
    \myVec{\hat{s}}(\myVec{r}) = f_{\myVec{\theta}}(\mathbf{u}(\myVec{r}))^T \mathbf{u}(\myVec{r}),
    \label{eqn:bf_ABLE}
    \end{equation}
    where $\hat{\myVec{s}}$ denotes the predicted beamformed signal. The trainable parameters here constitute the weights and biases of four fully-connected layers of which ABLE is comprised \cite{luijten2020adaptive}, along with the \ac{adc} parameters and the filters $\myMat{H}_1,\myMat{H}_2$. Finally, a B-mode image is obtained by envelope detection and logarithmic compression of the beamformed output.

    

    
    % ...
    
   
    \subsubsection{Training}
    Because of the non-uniform distribution of ultrasound data, we initialize the \acp{adc} with a logarithmic quantization rule (exponentially spaced), which is known to result in a higher quantization resolution in the low-intensity ranges \cite{Guilherme2003}.
    Training is based on  a signed-mean-squared-logarithmic-error loss function, defined as
    \begin{multline}
    \mathcal{L}(\myVec{\hat{s}},\myVec{s}) = \frac{1}{2}\lVert\log_{10}(\myVec{\hat{s}}^+)-\log_{10}(\myVec{s}^+)\rVert_2^2\quad+ \\
					\frac{1}{2}\lVert\log_{10}(-\myVec{\hat{s}}^-)-\log_{10}(-\myVec{s}^-)\rVert_2^2,
    \end{multline}
    where $\myVec{\hat{s}}$ and $\myVec{s}$ denote the predicted and target frames, respectively, and $(\cdot)^\pm$ denote the positive and negative signal components. Training the network parameters $f_{\myVec{\theta}}$ and $g_{\myVec{\phi}}$ can then be formulated as a regression problem, aiming at setting $\myVec{\theta}$ and $\myVec{\phi}$ to minimize the loss between a desired image $\myVec{s}$ and its reconstructed one obtained via \eqref{eqn:bf_ABLE}.
    
	
    \subsubsection{Evaluation}
    We train a set of models with different compression ratios by changing the rate of analog combining and the number quantization levels (bits) in the ADC. For each level of compression, numerical performance is assessed by measuring the contrast-to-noise ratio (CNR) over a simulated anechoic cyst phantom from the PICMUS dataset \cite{Liebgott2016picmus}. The CNR is defined as
    \begin{equation}
    \text{CNR} = 20\log_{10}\left(\frac{|\mu_{\rm low}-\mu_{\rm high}|}{\sqrt{(\sigma_{\rm low}^2+\sigma_{\rm high}^2)/2}}\right),
    \end{equation}
    where $\mu_{\rm low}$, $\mu_{\rm high}$, $\sigma_{\rm low}^2$ and $\sigma_{\rm high}^2$ represent the mean intensities and the variances of the anechoic and hyperechoic regions, respectively. Furthermore we evaluate the mean-absolute-error (MAE) between the baseline (uncompressed) and compressed reconstructions. It should be noted however, that this metric does not directly provide an measure of image quality, but gives a good indication of similarity to the training target. 
    	
    \begin{table}[]
    \caption{Numerical metrics}
    \centering
    \begin{tabular}{lllll}
    & Compression & CNR & MAE &  \\
    \hline
    \hline
    \multirow{3}{*}{Learned ADC} & 8x      & 7.01dB& 4.17 &  \\
    &  4x      & 10.39dB& 2.99 &  \\
    &  2.67x   & 12.25dB& 1.3 & \\
    \hline
    \multirow{3}{*}{Fixed ADC} & 8x      & 4.71dB& 4.28 &  \\
    &  4x      & 8.47dB& 3.62 &  \\
    &  2.67x   & 11.98dB& 1.33 & \\
    \hline
    \multirow{1}{*}{Baseline} & 1x      & 12.22dB& 0 &  \\

    \end{tabular}
    \label{table:US_results}
    \end{table}

    \vspace{-0.1cm}
    \subsection{Results}
    \label{subsec:AppBFResults}
    \vspace{-0.1cm}
    
    We evaluate deep task-based acquisition with the ABLE digital beamformer for different levels of data compression by varying the amount of analog combining and quantization in the model. To that end, we demonstrate $3$ settings, at compression rations of $2.6$, $4$ and $8$. These results are compared against a baseline model, which is ABLE without deep task-based acquisition\cite{luijten2020adaptive}. Furthermore we compare the learned ADC strategy with a deterministic (non-learned) approach having the same bit-budget, to show the performance difference between the two strategies.
    
    In Table \ref{table:US_results} we show the MAE and CNR for the different model settings. Additionally we provide the rate of analog combining and bitrate that achieve a specific compression ratio. As expected we see reduced CNR, and increased MAE compared to the target image, for higher compression ratios. Furthermore we see that in all cases, the task-based ADC framework outperforms the fixed scheme. That is, using a fixed logarithmic quantization rule, and ABLE as beamformer.
    
    To demonstrate that the performance metrics in  Table \ref{table:US_results} are indeed translated into a clear ultrasound image acquired in a compressed manner, we show in Fig.~\ref{fig:results_US}  the reconstructed images of two in-vivo records for the baseline model, the deep task-based \ac{adc}, and the target algorithm. Here, we can see that the model can handle reconstruction at tight bit-budgets well, yielding similar to baseline images. From the different compression levels we see that at rates of $2.6$x and even $4$x compression, the images are not notably affected in terms of image quality. However, as can be expected, at more extreme rates (i.e.. $8$x) the images start to get more blurry and fine details are lost. These results demonstrate the ability of deep task-based acquisition to facilitate operation with reduced number of bits in practical applications involving analog-to-digital conversion.
    

	

	\color{black}

	\vspace{-0.2cm}
	\section{Conclusions}
	\label{sec:Conclusions}
	\vspace{-0.1cm}
	In this work we designed a deep task-based acquisition system which learns to map a set of  analog signals into an estimate of an underlying task vector, obtained in the digital domain, in a data-driven manner. Our system adjusts its \ac{adc} mapping by approximating its continuous-to-discrete conversions using differentiable functions, 
	allowing to learn non-uniform mappings and to train the overall system in an end-to-end fashion. The proposed system was evaluated in both a synthetic detection setup as well as for ultrasound beamforming scenario, demonstrating its gains over using  uniform \acp{adc} and digital processing. 

    

	\bibliographystyle{IEEEtran}
	\bibliography{IEEEabrv,refs}

\begin{thebibliography}{10}
\providecommand{\url}[1]{#1}
\csname url@samestyle\endcsname
\providecommand{\newblock}{\relax}
\providecommand{\bibinfo}[2]{#2}
\providecommand{\BIBentrySTDinterwordspacing}{\spaceskip=0pt\relax}
\providecommand{\BIBentryALTinterwordstretchfactor}{4}
\providecommand{\BIBentryALTinterwordspacing}{\spaceskip=\fontdimen2\font plus
\BIBentryALTinterwordstretchfactor\fontdimen3\font minus
  \fontdimen4\font\relax}
\providecommand{\BIBforeignlanguage}[2]{{%
\expandafter\ifx\csname l@#1\endcsname\relax
\typeout{** WARNING: IEEEtran.bst: No hyphenation pattern has been}%
\typeout{** loaded for the language `#1'. Using the pattern for}%
\typeout{** the default language instead.}%
\else
\language=\csname l@#1\endcsname
\fi
#2}}
\providecommand{\BIBdecl}{\relax}
\BIBdecl

\bibitem{shlezinger2020learning}
N.~Shlezinger, R.~J. van Sloun, I.~A. Huijben, G.~Tsintsadze, and Y.~C. Eldar,
  ``Learning task-based analog-to-digital conversion for {MIMO} receivers,'' in
  \emph{Proc. IEEE ICASSP}, 2020, pp. 9125--9129.

\bibitem{eldar2015sampling}
Y.~C. Eldar, \emph{Sampling theory: Beyond bandlimited systems}.\hskip 1em plus
  0.5em minus 0.4em\relax Cambridge University Press, 2015.

\bibitem{walden1999analog}
R.~H. Walden, ``Analog-to-digital converter survey and analysis,'' \emph{{IEEE}
  J. Sel. Areas Commun.}, vol.~17, no.~4, pp. 539--550, 1999.

\bibitem{kipnis2018analog}
A.~Kipnis, Y.~C. Eldar, and A.~J. Goldsmith, ``Analog-to-digital compression: A
  new paradigm for converting signals to bits,'' \emph{{IEEE} Signal Process.
  Mag.}, vol.~35, no.~3, pp. 16--39, 2018.

\bibitem{xiao2017millimeter}
M.~Xiao, S.~Mumtaz, Y.~Huang, L.~Dai, Y.~Li, M.~Matthaiou, G.~K. Karagiannidis,
  E.~Bj{\"o}rnson, K.~Yang, and I.~Chih-Lin, ``Millimeter wave communications
  for future mobile networks,'' \emph{{IEEE} J. Sel. Areas Commun.}, vol.~35,
  no.~9, pp. 1909--1935, 2017.

\bibitem{chernyakova2014fourier}
T.~Chernyakova and Y.~C. Eldar, ``Fourier-domain beamforming: the path to
  compressed ultrasound imaging,'' \emph{{IEEE} Trans. Ultrason., Ferroelectr.,
  Freq. Control}, vol.~61, no.~8, pp. 1252--1267, 2014.

\bibitem{yazicigil2019taking}
R.~T. Yazicigil, T.~Haque, P.~R. Kinget, and J.~Wright, ``Taking compressive
  sensing to the hardware level: Breaking fundamental radio-frequency hardware
  performance tradeoffs,'' \emph{{IEEE} Signal Process. Mag.}, vol.~36, no.~2,
  pp. 81--100, 2019.

\bibitem{jain2020esampling}
N.~Jain, N.~Shlezinger, B.~Tiwari, Y.~C. Eldar, A.~Gupta, V.~A. Bohara, and
  P.~G. Bahubalindruni, ``esampling: Energy harvesting {ADCs},'' \emph{arXiv
  preprint arXiv:2007.08275}, 2020.

\bibitem{6936944}
S.~{Lee}, A.~P. {Chandrakasan}, and H.~{Lee}, ``A 1 {GS/s 10b 18.9 mW}
  time-interleaved {SAR ADC} with background timing skew calibration,''
  \emph{{IEEE} J. Solid-State Circuits}, vol.~49, no.~12, pp. 2846--2856, Dec
  2014.

\bibitem{8727467}
Y.~{Zhou}, B.~{Xu}, and Y.~{Chiu}, ``A 12-b {1-GS/s 31.5-mW} time-interleaved
  {SAR ADC} with analog {HPF}-assisted skew calibration and randomly sampling
  reference {ADC},'' \emph{{IEEE} J. Solid-State Circuits}, vol.~54, no.~8, pp.
  2207--2218, Aug 2019.

\bibitem{shlezinger2018hardware}
N.~Shlezinger, Y.~C. Eldar, and M.~R. Rodrigues, ``Hardware-limited task-based
  quantization,'' \emph{{IEEE} Trans. Signal Process.}, vol.~67, no.~20, pp.
  5223--5238, 2019.

\bibitem{shlezinger2018asymptotic}
------, ``Asymptotic task-based quantization with application to massive
  {MIMO},'' \emph{{IEEE} Trans. Signal Process.}, vol.~67, no.~15, pp.
  3995--4012, 2019.

\bibitem{Salamtian19task}
S.~Salamtian, N.~Shlezinger, Y.~C. Eldar, and M.~Medard, ``Task-based
  quantization for recovering quadratic functions using principal inertia
  components,'' in \emph{Proc. IEEE ISIT}, 2019.

\bibitem{shlezinger2019deep}
N.~Shlezinger and Y.~C. Eldar, ``Deep task-based quantization,''
  \emph{Entropy}, vol.~23, no.~1, p. 104, 2021.

\bibitem{huijben2019learning}
I.~A.~M. Huijben, B.~S. Veeling, K.~Janse, M.~Mischi, and R.~J.~G. van Sloun,
  ``Learning sub-sampling and signal recovery with applications in ultrasound
  imaging,'' \emph{{IEEE} Trans. Med. Imag.}, vol.~39, no.~12, pp. 3955--3966,
  2020.

\bibitem{mulleti2021learning}
S.~Mulleti, H.~Zhang, and Y.~C. Eldar, ``Learning to sample: Data-driven
  sampling and reconstruction of {FRI} signals,'' \emph{arXiv preprint
  arXiv:2106.14500}, 2021.

\bibitem{solodky2018sampling}
G.~Solodky and M.~Feder, ``Sampling a noisy multiple output channel to maximize
  the capacity,'' in \emph{Proc. IEEE EUSIPCO}, 2018, pp. 445--449.

\bibitem{liu2018analog}
X.~Liu, E.~G{\"o}n{\"u}lta{\c{s}}, and C.~Studer, ``Analog-to-feature ({A2F})
  conversion for audio-event classification,'' in \emph{Proc. IEEE EUSIPCO)},
  2018.

\bibitem{neuhaus2020task}
P.~Neuhaus, N.~Shlezinger, M.~D{\"o}rpinghaus, Y.~C. Eldar, and G.~Fettweis,
  ``Task-based analog-to-digital converters,'' \emph{{IEEE} Trans. Signal
  Process.}, vol.~69, pp. 5403--5418, 2021.

\bibitem{kipnis2016distortion}
A.~Kipnis, A.~J. Goldsmith, Y.~C. Eldar, and T.~Weissman, ``Distortion rate
  function of sub-nyquist sampled gaussian sources,'' \emph{{IEEE} Trans. Inf.
  Theory}, vol.~62, no.~1, pp. 401--429, 2016.

\bibitem{kipnis2018fundamental}
A.~Kipnis, Y.~C. Eldar, and A.~J. Goldsmith, ``Fundamental distortion limits of
  analog-to-digital compression,'' \emph{{IEEE} Trans. Inf. Theory}, vol.~64,
  no.~9, pp. 6013--6033, 2018.

\bibitem{agustsson2017soft}
E.~Agustsson, F.~Mentzer, M.~Tschannen, L.~Cavigelli, R.~Timofte, L.~Benini,
  and L.~V. Gool, ``Soft-to-hard vector quantization for end-to-end learning
  compressible representations,'' in \emph{Advances in Neural Information
  Processing Systems}, 2017, pp. 1141--1151.

\bibitem{frazier2018tutorial}
P.~I. Frazier, ``A tutorial on {Bayesian} optimization,'' \emph{arXiv preprint
  arXiv:1807.02811}, 2018.

\bibitem{shlezinger19joint}
N.~Shlezinger, S.~Salamtian, Y.~C. Eldar, and M.~Medard, ``Joint sampling and
  recovery of correlated sources,'' in \emph{Proc. IEEE ISIT}, 2019.

\bibitem{mendez2016hybrid}
R.~M{\'e}ndez-Rial, C.~Rusu, N.~Gonz{\'a}lez-Prelcic, A.~Alkhateeb, and R.~W.
  Heath, ``Hybrid {MIMO} architectures for millimeter wave communications:
  Phase shifters or switches?'' \emph{{IEEE} Access}, vol.~4, pp. 247--267,
  2016.

\bibitem{ioushua2019family}
S.~S. Ioushua and Y.~C. Eldar, ``A family of hybrid analog--digital beamforming
  methods for massive {MIMO} systems,'' \emph{{IEEE} Trans. Signal Process.},
  vol.~67, no.~12, pp. 3243--3257, 2019.

\bibitem{gong2019rf}
T.~Gong, N.~Shlezinger, S.~S. Ioushua, M.~Namer, Z.~Yang, and Y.~C. Eldar,
  ``{RF} chain reduction for {MIMO} systems: A hardware prototype,''
  \emph{{IEEE} Syst. J.}, vol.~14, no.~4, pp. 5296--5307, 2020.

\bibitem{rose1992vector}
K.~Rose, E.~Gurewitz, and G.~C. Fox, ``Vector quantization by deterministic
  annealing,'' \emph{{IEEE} Trans. Inf. Theory}, vol.~38, no.~4, pp.
  1249--1257, 1992.

\bibitem{shlezinger2019dynamic}
N.~Shlezinger, O.~Dicker, Y.~C. Eldar, I.~Yoo, M.~F. Imani, and D.~R. Smith,
  ``Dynamic metasurface antennas for uplink massive {MIMO} systems,''
  \emph{{IEEE} Trans. Commun.}, vol.~67, no.~10, pp. 6829--6843, 2019.

\bibitem{wang2020dynamic}
H.~Wang, N.~Shlezinger, Y.~C. Eldar, S.~Jin, M.~F. Imani, I.~Yoo, and D.~R.
  Smith, ``Dynamic metasurface antennas for {MIMO-OFDM} receivers with
  bit-limited {ADCs},'' \emph{{IEEE} Trans. Commun.}, vol.~69, no.~4, pp.
  2643--2659, 2020.

\bibitem{shlezinger2020dynamic}
N.~Shlezinger, G.~C. Alexandropoulos, M.~F. Imani, Y.~C. Eldar, and D.~R.
  Smith, ``Dynamic metasurface antennas for {6G} extreme massive {MIMO}
  communications,'' \emph{{IEEE} Wireless Commun.}, vol.~28, no.~2, pp.
  106--113, 2021.

\bibitem{mead1990neuromorphic}
C.~Mead, ``Neuromorphic electronic systems,'' \emph{Proc. {IEEE}}, vol.~78,
  no.~10, pp. 1629--1636, 1990.

\bibitem{danial2018breaking}
L.~Danial, N.~Wainstein, S.~Kraus, and S.~Kvatinsky, ``Breaking through the
  speed-power-accuracy tradeoff in {ADCs} using a memristive neuromorphic
  architecture,'' \emph{IEEE Trans. Emerg. Topics Comput. Intell.}, vol.~2,
  no.~5, pp. 396--409, 2018.

\bibitem{vinyals2016matching}
O.~Vinyals, C.~Blundell, T.~Lillicrap, D.~Wierstra \emph{et~al.}, ``Matching
  networks for one shot learning,'' in \emph{Advances in neural information
  processing systems}, 2016, pp. 3630--3638.

\bibitem{maclaurin2015gradient}
D.~Maclaurin, D.~Duvenaud, and R.~Adams, ``Gradient-based hyperparameter
  optimization through reversible learning,'' in \emph{International Conference
  on Machine Learning}, 2015, pp. 2113--2122.

\bibitem{wichrowska2017learned}
O.~Wichrowska, N.~Maheswaranathan, M.~W. Hoffman, S.~G. Colmenarejo, M.~Denil,
  N.~de~Freitas, and J.~Sohl-Dickstein, ``Learned optimizers that scale and
  generalize,'' in \emph{Proceedings of the 34th International Conference on
  Machine Learning-Volume 70}.\hskip 1em plus 0.5em minus 0.4em\relax JMLR.
  org, 2017, pp. 3751--3760.

\bibitem{finn2017model}
C.~Finn, P.~Abbeel, and S.~Levine, ``Model-agnostic meta-learning for fast
  adaptation of deep networks,'' \emph{arXiv preprint arXiv:1703.03400}, 2017.

\bibitem{brochu2010tutorial}
E.~Brochu, V.~M. Cora, and N.~De~Freitas, ``A tutorial on {Bayesian}
  optimization of expensive cost functions, with application to active user
  modeling and hierarchical reinforcement learning,'' \emph{arXiv preprint
  arXiv:1012.2599}, 2010.

\bibitem{Balandat2019botorch}
M.~Balandat, B.~Karrer, D.~R. Jiang, S.~Daulton, B.~Letham, A.~G. Wilson, and
  E.~Bakshy, ``Botorch: Programmable bayesian optimization in pytorch,''
  \emph{arXiv preprint arXiv:1910.06403}, 2019.

\bibitem{ax}
\BIBentryALTinterwordspacing
``Ax adaptive experimentation platform.'' [Online]. Available:
  \url{https://ax.dev/}
\BIBentrySTDinterwordspacing

\bibitem{kingma2014adam}
D.~P. Kingma and J.~Ba, ``Adam: A method for stochastic optimization,''
  \emph{arXiv preprint arXiv:1412.6980}, 2014.

\bibitem{luijten2020adaptive}
B.~Luijten, R.~Cohen, F.~J. De~Bruijn, H.~A. Schmeitz, M.~Mischi, Y.~C. Eldar,
  and R.~J. Van~Sloun, ``Adaptive ultrasound beamforming using deep learning,''
  \emph{{IEEE} Trans. Med. Imag.}, vol.~39, no.~12, pp. 3967--3978, 2020.

\bibitem{Guilherme2003}
J.~Guilherme and J.~Vital, \emph{Logarithmic Analogue-to-Digital Converters},
  01 2003, pp. 241--275.

\bibitem{Liebgott2016picmus}
H.~Liebgott, A.~Rodriguez-Molares, F.~Cervenansky, J.~A. Jensen, and
  O.~Bernard, ``Plane-wave imaging challenge in medical ultrasound,'' in
  \emph{2016 IEEE International Ultrasonics Symposium (IUS)}, Sept. 2016.

\end{thebibliography}

\end{document}